\documentclass[11pt]{article}
\usepackage{amsmath}
\usepackage{amssymb}
\usepackage{amsthm}
\usepackage{bigfoot}
\usepackage{booktabs}
\usepackage{etoolbox}
\usepackage{float}
\usepackage[margin=1in]{geometry}
\usepackage{graphicx}
\usepackage{mathrsfs}
\usepackage{natbib}
\usepackage{xcolor}
\usepackage[hidelinks]{hyperref}

\DeclareNewFootnote{AAffil}[arabic]
\DeclareNewFootnote{ANote}[fnsymbol]

\makeatletter
\patchcmd\maketitle{\def\@makefnmark{\rlap{\@textsuperscript{\normalfont\@thefnmark}}}}{}{}{}
\makeatother

\makeatletter
\def\thanksAAffil#1{%
  \footnotemarkAAffil\protected@xdef\@thanks{\@thanks%
        \protect\footnotetextAAffil[\the \c@footnoteAAffil]{#1}}%
}
\def\thanksANote#1{%
  \footnotemarkANote%
  \protected@xdef\@thanks{\@thanks%
        \protect\footnotetextANote[\the \c@footnoteANote]{#1}}%
}
\makeatother

\providecommand{\keywords}[1]{\small \textbf{\textit{Keywords:}} #1}
\newtheorem{assumption}{Assumption}
\newtheorem{proposition}{Proposition}
\graphicspath{{figs/}}

\title{Deep Learning for Solving and Estimating Dynamic Macro-Finance Models}

\author{
  Benjamin Fan\thanksAAffil{Department of Mathematics, Massachusetts Institute of Technology, Cambridge, MA 02139, USA}$^{,}$\thanksANote{These authors contributed equally to this work.},
  Edward Qiao\footnotemarkAAffil[1]$^{,}$\footnotemarkANote[1],
  Anran Jiao\thanksAAffil{Department of Computer and Information Science, University of Pennsylvania, Philadelphia, PA 19104, USA},
  Zhouzhou Gu\thanksAAffil{Department of Economics, Princeton University, Princeton, NJ 08544, USA},
  Wenhao Li\thanksAAffil{Marshall School of Business, University of Southern California, Los Angeles, CA 90089, USA},
  Lu Lu\thanksAAffil{Department of Chemical and Biomolecular Engineering, University of Pennsylvania, Philadelphia, PA 19104, USA}$^{,}$\thanksANote{Corresponding author. Email: lulu1@seas.upenn.edu}
}

\date{}

\begin{document}

\maketitle

\begin{abstract}
We develop a methodology that utilizes deep learning to simultaneously solve and estimate canonical continuous-time general equilibrium models in financial economics. We illustrate our method in two examples: (1) industrial dynamics of firms and (2) macroeconomic models with financial frictions. Through these applications, we illustrate the advantages of our method: generality, simultaneous solution and estimation, leveraging the state-of-art machine-learning techniques, and handling large state space. The method is versatile and can be applied to a vast variety of problems.
\end{abstract}


\section{Introduction}

Dynamic equilibrium models are the cornerstones of the fast-growing macro-finance literature that tries to understand how financial frictions and asset prices influence economic dynamics, in addition to addressing important policy questions including the design and impact of financial regulation, industrial policy, and monetary policy \citep{he2013intermediary, brunnermeier2014macroeconomic, drechsler2014model, bianchi2014banks, gertler2015banking, huang2018banking,hansen2018comparative,di2019optimal, li2019public, krishnamurthy2020dissecting, maxted2020macro}. These models feature high degrees of nonlinearity originating from either agents' financial constraints or preferences, which make the linearization methods widely used in the macro literature infeasible.

The literature has thus far mostly focused on highly tractable models with a small number of state variables (typically one or two). Furthermore, since solving these models numerically, such as by finite differences~\citep{achdou2014partial,brunnermeier2014macroeconomic}, could be quite time-consuming, model parameters are often picked by calibration, which involves intensive model evaluation. Matching moments involves solving the model, simulating the model for a long period and calculating the moment value, and repeating the same procedure for a large number of parameter combinations. Although simulated methods of moment have been applied to corporate-finance models \citep{gomes2003equilibrium, whited2006financial, hennessy2007costly, matvos2014resource}, dynamic equilibrium models are restricted by the curse of dimensionality. Additionally, taking expectations is typical in dynamic problems, but it incurs a significant computational burden. Finally, for different problems, researchers typically need to tailor their numerical methods, which limits the accessibility of the literature, and these methods do not automatically take advantage of the rapidly evolving computational tools.

Machine learning (ML) models have recently been used in economics and finance, but mainly for the purpose of better predicting economic and financial outcomes in markets, such as stocks, insurance, corporate bankruptcy, and cryptocurrency. Recent research papers \citep{fernandez2020solving, chen2021deep, han2021deepham, duarte2018machine, azinovic2022deep, maliar2021deep, gopalakrishna2020macro,huang2022probabilistic} have utilized deep learning techniques to solve economic models. Many of these studies leverage deep neural networks to approximate value functions and policy functions, and often combine reinforcement learning techniques. While some of these studies have focused on discrete-time models \citep{han2021deepham,maliar2021deep,azinovic2022deep}, our paper aims to apply deep learning methods to continuous-time models \citep{fernandez2020solving,duarte2018machine,gopalakrishna2020macro,huang2022probabilistic}, which have been widely used in financial economics and allow for more tractability \citep{leland1994corporate}. Specifically, we address the control problem and parameter estimation by jointly solving the Hamilton-Jacobi-Bellman Equation, Kolmogorov Forward Equation, and moment conditions.

We contribute to this growing research area by applying a recently-developed deep-learning method for solving partial differential equations (PDEs), physics-informed neural networks (PINNs) \citep{raissi2019pinn,lu2021deepxde,karniadakis2021physics,wu2023comprehensive}, to economic settings, simultaneously solving and estimating model parameters and allowing for heterogeneous agents. PINNs work by embedding PDE residuals into the loss function of the neural network via automatic differentiation \citep{lu2021deepxde,yu2021gradient}. As such, approximating the solution of PDEs is no more than minimizing the loss function, which can be done with gradient descent techniques. This method of solving PDEs is mesh-free and easy to implement, and it can be applied to a wide variety of PDE types. In addition to solving such forward problems, PINNs can also be used to solve inverse PDE problems, which involve predicting the values of unknown parameters given a set of measurements or observations of the solution. PINNs have achieved success with both forward and inverse problems in a diverse range of fields, including optics \citep{chen2020physics,lu2021physics}, systems biology \citep{yazdani2020systems,daneker2023systems}, fluid mechanics \citep{raissi2020fluid, tartakovsky2020fluid}, solid mechanics \citep{wu2022effective}, biomedicine \citep{kissas2020biomed, costabal2020biomed}, and other types of PDEs \citep{pang2019fpinns,zhang2019quantifying}. However, there have been fewer applications to problems in economics, which this paper will explore.

In this study, we consider two models. First, we solve a model of industrial dynamics with financial frictions, which considers an industrial equilibrium of a banking sector that takes deposits, makes loans, and uses labor input to manage deposits and loans. Then, we consider a macroeconomic model with the financial sector featuring binding constraints, non-linear financial amplifications, and boundary singularity. We demonstrate the advantage of our methodology in providing a general framework for solving PDEs, as opposed to traditional methods, which require designing different algorithms for different problems. We also display the ease of solving inverse problems, in which we assume parameters are unknown and impose moment conditions. Thus, we are able to simultaneously solve and estimate parameters.

Our methods have the following advantages: (1) the model is solved globally; (2) the method allows for higher dimensionality; (3) deep learning is a proven and effective method in solving PDEs; (4) parameter estimation and PDE solutions are obtained simultaneously; (5) differentiation is handled analytically without numerical discretization error;  (6) the underlying package constantly integrates the state-of-art deep-learning algorithms, so the method keeps improving itself; and (7) the method is versatile and can be applied to a vast variety of problems.

This paper is organized as follows. In Section~\ref{sec:pde intro}, we introduce the PINN method for solving both forward and inverse PDEs. Then, in Sections~\ref{sec:firm dynamic} and \ref{sec:brunnermeier sannikov}, we apply the PINN method to solve several problems in economics. Section~\ref{sec:conclusion} summarizes the findings and concludes the paper.
\section{Deep learning for PDEs}
\label{sec:pde intro}

We first describe general deep neural networks, allowing us to present the framework of physics-informed neural networks (PINNs), which will be leveraged to solve forward and inverse problems (i.e., model solution and model estimation) for PDEs. 
Afterward, we provide an overview of the hyperparameters used in the following sections.

\subsection{Deep neural networks}
\label{sec:fnn info}

\begin{figure}[htbp]
    \centering
    \includegraphics[scale=0.25]{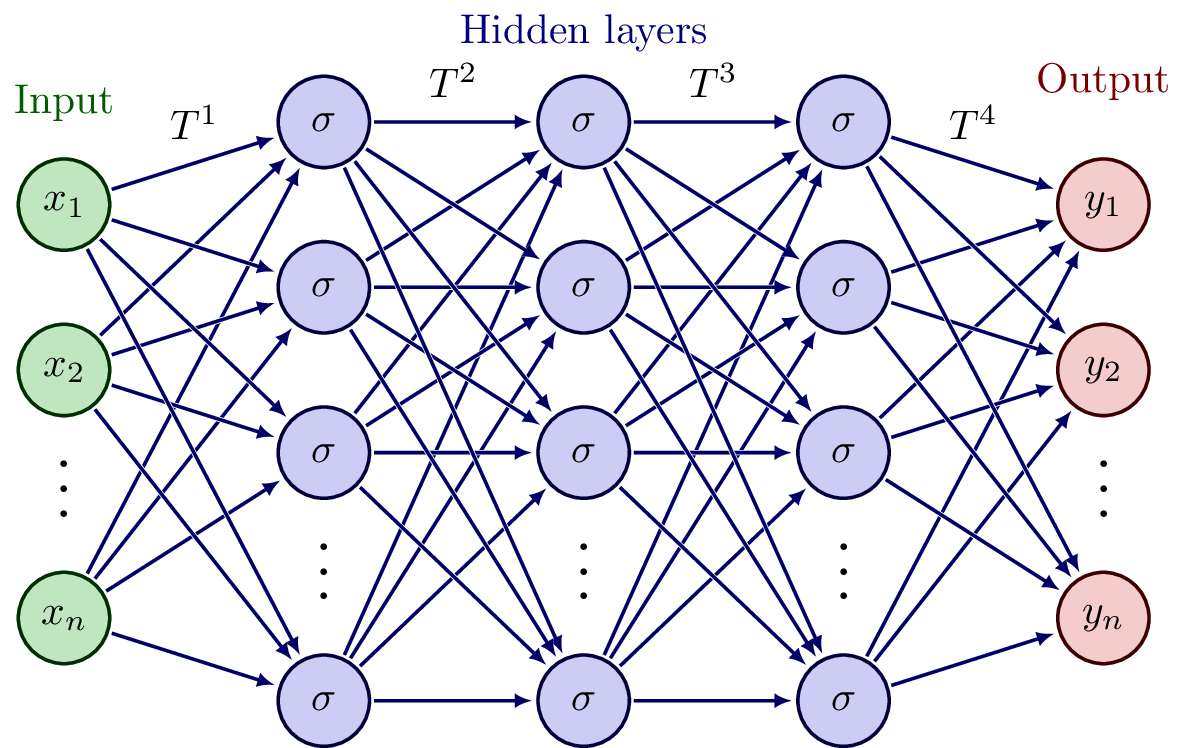}
    \caption{\textbf{Visualization of a deep neural network.} In this example, the number of layers $L$ is~$4$.}
    \label{fig:fnn}
\end{figure}

Although there are several different types of deep neural networks that can be used in PINNs, throughout this paper, we use the classical feed-forward neural networks (FNNs) as our network architecture. An $L$-layer FNN is a function $\mathcal{N}^L(\mathbf{x})\colon\mathbb{R}^{d_{\text{in}}} \to \mathbb{R}^{d_\text{out}}$ with $L-1$ hidden layers such that the $\ell$-th layer has $N_\ell$ neurons. Clearly, $N_0 = d_{\text{in}}$ and $N_L = d_{\text{out}}$, where $d_{\text{in}}$ is the input dimension and $d_{\text{out}}$ is the output dimension. Furthermore, for each $1 \le \ell \le L$, we define a weight matrix $\mathbf{W}^\ell \in \mathbb{R}^{N_\ell \times N_{\ell-1}}$ and bias vector $\mathbf{b}^\ell \in \mathbb{R}^{N_\ell}$. Then, letting $T^\ell(\mathbf{x}) = \mathbf{W}^\ell \mathbf{x} + \mathbf{b}^\ell$ be the affine transformation in the $\ell$-th layer, for some non-linear activation function $\sigma$, we have
\[\mathcal{N}^L(\mathbf{x}) = T^L \circ \sigma \circ T^{L-1} \circ \ldots \circ \sigma \circ T^1(\mathbf{x}).\]
A network with $L = 4$ is visualized in Fig.~\ref{fig:fnn}. There are different possible activation functions $\sigma$, and in this study, we either use the hyperbolic tangent ($\tanh$) or the Swish \citep{ramachandran2017swish} activation function given by $\mathrm{Swish}(x)= \frac{x}{1+e^{-x}}$. The former is a typical activation function in deep learning, while the latter can better deal with problems with steep gradients, which are features of the problems we will discuss.

\subsection{PINNs for solving forward PDEs}

We first discuss the use of PINNs to solve forward problems of PDEs. Consider the following PDE parameterized by $\boldsymbol\lambda$ with solution $u(\mathbf{x}, t)$ for $\mathbf{x} = (x_1, \dots, x_d)$ over the domain $\Omega \subset \mathbb{R}^d$:
\[f\left(\mathbf{x}; \frac{\partial u}{\partial x_1}, \dots, \frac{\partial u}{\partial x_d}; \frac{\partial^2 u}{\partial x_1 \partial x_1}, \dots, \frac{\partial^2 u}{\partial x_1\partial x_d}; \dots ; \boldsymbol\lambda \right) = 0, \quad \mathbf{x} \in \Omega \tag{2.1} \label{eqn:2.1}\]
with boundary conditions
\[\mathcal{B}(u, \mathbf{x}) = 0 \quad \text{on} \quad \partial \Omega.\]

To find the solution, we build a neural network $\hat{u}(\mathbf{x}; \boldsymbol\theta)$ with trainable parameters $\boldsymbol\theta = \{ \mathbf{W}^\ell, \mathbf{b}^\ell \}_{\ell=1}^L$. To train the network, we use $\mathcal{T}_f$ points inside the domain and $\mathcal{T}_b$ points on the boundary. Then, the loss function is defined as 
\[ \mathcal{L}(\boldsymbol\theta; \mathcal{T}) = w_f\mathcal{L}_f(\boldsymbol\theta; \mathcal{T}_f) + w_b\mathcal{L}_b(\boldsymbol\theta; \mathcal{T}_b) \tag{2.2} \label{eqn:2.2}\] 
with 
\begin{align*}
    \mathcal{L}_f(\boldsymbol\theta; \mathcal{T}_f) &= \frac{1}{|\mathcal{T}_f|} \sum_{\mathbf{x} \in \mathcal{T}_f} \left\Vert f\left(\mathbf{x}; \frac{\partial \hat{u}}{\partial x_1}, \dots, \frac{\partial \hat{u}}{\partial x_d}; \frac{\partial^2 \hat{u}}{\partial x_1 \partial x_1}, \dots, \frac{\partial^2 \hat{u}}{\partial x_1\partial x_d}; \dots ; \boldsymbol\lambda \right)\right\Vert^2_2, \\
    \mathcal{L}_b(\boldsymbol\theta, \mathcal{T}_b) &= \frac{1}{|\mathcal{T}_b|} \sum_{\mathbf{x} \in \mathcal{T}_b} \Vert\mathcal{B}(\hat{u}, \mathbf{x})\Vert^2_2,
\end{align*}
and $w_f$ and $w_b$ are the weights. After we have the loss function, we can train the neural network by minimizing $\mathcal{L}(\boldsymbol\theta; \mathcal{T})$ using gradient-based optimizers, such as Adam \citep{kingma2015adam}, i.e,
$$\boldsymbol{\theta}^{*} = \arg{\min_{\boldsymbol{\theta}}\mathcal{L}(\boldsymbol{\theta}; \mathcal{T})}$$
and then obtain an approximated solution $\hat{u}(\mathbf{x}; \boldsymbol{\theta}^{*})$. For boundary conditions, we note that instead  of using a loss function, it is often possible to enforce boundary conditions automatically and exactly by constructing a special form of the approximated solution \citep{lu2021physics}. The form is problem dependent, and we will discuss the details in the section of each problem.

\subsection{PINNs for solving inverse PDEs}

Next, we discuss using PINNs to solve inverse problems for PDEs \citep{raissi2019pinn, lu2021deepxde}. Inverse problems involve some unknown parameters $\boldsymbol{\lambda}$ in Eq.~\eqref{eqn:2.1} to be solved for, but we are given some extra information besides the PDE and boundary conditions:
\[\mathcal{I}(u, \mathbf{x}) = 0.\]
For the problems considered in this paper, this extra information comes in the form of moment targets.

Training the PINN for the inverse problem is almost identical as the forward problem, except that the loss function in Eq.~\eqref{eqn:2.2} has an extra loss term:
\[\mathcal{L}(\boldsymbol{\theta}, \boldsymbol{\lambda}; \mathcal{T}) = w_f\mathcal{L}_f(\boldsymbol{\theta}, \boldsymbol{\lambda}; \mathcal{T}_f) + w_b\mathcal{L}_b(\boldsymbol\theta, \boldsymbol{\lambda}; \mathcal{T}_b) + w_i\mathcal{L}_i(\boldsymbol\theta, \boldsymbol{\lambda}),\]
where
\[\mathcal{L}_i(\boldsymbol{\theta}, \boldsymbol{\lambda}) = \Vert\mathcal{I}(\hat{u}, \mathbf{x})\Vert^2_2,\]
is the error of the additional information $\mathcal{I}$. When solving forward problems, we only optimize $\boldsymbol{\theta}$, and for inverse problems, we optimize both $\boldsymbol{\theta}$ and $\boldsymbol{\lambda}$ together, i.e., our solution is
$$\boldsymbol{\theta}^{*}, \boldsymbol{\lambda}^{*} = \arg{\min_{\boldsymbol{\theta}, \boldsymbol{\lambda}}\mathcal{L}(\boldsymbol{\theta}, \boldsymbol{\lambda}; \mathcal{T})}.$$
This approach incorporates model solution (i.e., solving for the solution $u$) and model estimation (i.e., solving for the parameters $\boldsymbol{\lambda}$) together in a consistent way.  Furthermore, we can easily generalize this method to a PDE system with multiple functions to be solved and multiple terms of extra information.

\subsection{Implementation}

We apply PINNs to solve several forward and inverse PDE problems in economics. To improve the training of the network, we use a learning rate decay with the formula
\[\gamma_n = \frac{\gamma_0}{1 + n/S},\]
where $n$ is the number of iterations, $\gamma_n$ is the learning rate after $n$ iterations, $\gamma_0$ is the initial learning rate, and $S$ is the decay step. Throughout the paper, we use the Adam optimizer. We list most hyperparameters in Table~\ref{tab:hyperparameters}, and more details can be found in the section of each problem. We implement the code with the Python library DeepXDE \citep{lu2021deepxde}, and the code used is publicly available on the GitHub repository \url{https://github.com/lu-group/pinn-macro-finance}.

\begin{table}[htbp]
    \centering
    \caption{\textbf{Hyperparameters used for each problem.}}
    \label{tab:hyperparameters}
    \begin{tabular}{l|cccccc}
        \toprule
        Section & Depth & Width & Activation & Learning rate $\gamma_0$ & Decay step $S$ & No.\ of iterations \\
        \midrule
        \ref{sec:hjb-forward} & 7 & 64 & tanh & $5 \times 10^{-4}$ & 6000 & $7.5 \times 10^4$ \\
        \ref{sec:hjb-inv-two-var} & 7 & 64 & tanh & $1 \times 10^{-3}$ & 2500 & $1.5 \times 10^5$ \\
        \ref{sec:hjb-inv-three-var} & 7 & 64 & tanh & $1 \times 10^{-3}$ & 2500 & $1.5 \times 10^5$ \\
        \ref{sec:intermediary-solution} & 7 & 128 & Swish & See Sec.~\ref{sec:brunnermeier-sannikov-tech-detail} & See Sec.~\ref{sec:brunnermeier-sannikov-tech-detail} & $3.0 \times 10^5$ \\
        \ref{sec:iap-one-param} & 6 & 64 & Swish & $1 \times 10^{-4}$ & 2000 & $2.0 \times 10^5$ \\
        \ref{sec:iap-two-params} & 6 & 64 & Swish & $5 \times 10^{-4}$ & 1500 & $1.5 \times 10^5$ \\
        \bottomrule
    \end{tabular}
\end{table}

\section{A model of industrial dynamics with financial frictions}
\label{sec:firm dynamic}

In this section, we consider an industrial equilibrium of a banking sector that takes deposits, makes loans, and uses labor input to manage deposits and loans. Labor productivity determines the number of loans and deposits that each bank can take. Both loan rate and deposit rates are endogenously determined via the competitive market equilibrium that features bank entry, exit, and equity payouts. The challenges of this problem are twofold: first, banks endogenously determine whether or not to enter or exit the market, so the problem features endogenous entry and exit boundaries; second, we need to track the entire distribution of banks in order to clear the market.

\subsection{Problem setup}
\label{sec:firm dynamic setup}

Time evolves continuously. All bank assets and debts are modeled as short-term. Banks can borrow via deposits at a rate $r^{d}$ and via capital market at a rate $r$ (think of this as the policy rate, e.g., FFR). Banks can lend at loan rate $r^{l}$ or the capital market rate $r$. Banks need to hire workers to serve their deposits and loans. For banks with productivity $z$, the number of loans that $l$ units of labor can serve is 
\[
f(z,l)=zl^{\alpha},\quad\alpha\in(0,1),
\]
which features decreasing return to scale. A rationale for this assumption is that as banks get bigger, it is increasingly difficult to find new depositors and new borrowers on which they can earn a profit. Similarly, the amount of deposits that $d$ units of labor can serve is 
\[
f(z,d)=zd^{\alpha}.
\]
Because banks can borrow and lend freely in the capital market at a rate $r$, they are not constrained by lending and deposit-taking choices, as long as they are nonnegative. 

The stochastic process for $z_t$ is given exogenously as 
\[
dz_{t}=\mu(z_{t})dt+\sigma(z_{t})dB_{t}, \]
with two reflecting boundaries $\underline{z}$ and $\bar{z}$. Denote bank equity as $e$. We impose the financial friction by
\[
f(z,l)\leq\phi e,
\]
\[
f(z,d)\leq\phi e.
\]
Banks incur a fixed operating cost $c_{f}$, and thus the instantaneous profit is
\[
\pi(e_{t},z_{t},l_{t},d_{t}) = \underbrace{r_{t}^{l}z_{t}l_{t}^{\alpha}}_{\text{lending revenue}} - \underbrace{r_{t}^{d}z_{t}d_{t}^{\alpha}}_{\text{deposits interest expense}} + \underbrace{(z_{t}d_{t}^{\alpha}+e_{t}-z_{t}l_{t}^{\alpha})r_{t}}_{\text{net capital market lending}} - \underbrace{w_{t}\cdot(l_{t}+d_{t})}_{\text{labor cost}} - \underbrace{c_{f}}_{\text{fixed cost}}.
\]

To simplify the problem, we assume full symmetry between the deposit market and the loan market. Both loan demand and deposit demand functions are in the same form:
\[r^{l}-r=\beta(L+L_{0})^{-\varepsilon},\]
\[r-r^{d}=\beta(D+D_{0})^{-\varepsilon},\]
where $D$ and $L$ are the aggregate amounts of deposits and loans, respectively. We assume $L_0 = D_0$ for simplicity. Due to the full symmetry, the loan spread is equal to the deposit spread, $r^l - r = r - r^d$. In what follows, we will only use the notations on the loan side.

Since the bank can freely adjust its labor input and scales of operations at each instance, the optimal decisions are
\[l^{*}=d^{*}=\min\left\{\left(\frac{(r^{l}-r)z\alpha}{w}\right)^{\frac{1}{1-\alpha}}, \left(\frac{\phi e}{z}\right)^{1/\alpha}\right\}.\]
Therefore, the optimized profit function is
$$\pi^{*}(e, z)=2\left(r^{l}-r\right) z\left(l^{*}\right)^{\alpha}+e \cdot r-2 w l^{*}-c_{f}.$$

Denote $v(e, z)$ as the value function for a bank with equity $e$ and productivity $z$, with $\underline{v}(e)=e$ the reservation value if a bank exists. We assume that banks pay out equity when they are financially unconstrained, and the equity payout function is
\[\zeta(e,z)= \max\left\{\kappa\left(\phi e-f(z,l^{*})\right), 0\right\}. \]
In other words, equity payout smoothly increases as the bank gets further away from the financial constraint. Then we can write the bank equity dynamics as 
\[de_{t}=\underbrace{\left(\pi^{*}(e_{t},z_{t})-1_{(e_{t},z_{t})\in\mathcal{C}^{c}}\cdot\zeta(e_{t},z_{z})\right)}_{\equiv\mu_{e}(e_{t},z_{t})}dt-1_{v(e_{t},z_{t})<\underline{v}(e_{t})}\cdot e_{t} ,\]
where the last term reflects the immediate exit when the bank continuation value is smaller than the liquidation value.

Banks optimize the expected discounted cash flows at the rate $r$. Then we have the Hamilton-Jacobi-Bellman (HJB) equation
$$r\cdot v(e,z)=\max\left\{ \pi^{*}(e,z)(1+v_{e}^{\prime})+(1-v_{e}^{\prime})\zeta(e,z)1_{(e,z)\in\mathscr{C}^{c}}+v_{z}^{\prime}\mu(z)+\frac{1}{2}v_{zz}^{\prime\prime}\sigma(z)^{2},r\underline{v}(e)\right\},$$
where the set $\mathscr{C}$ is the region where the bank is financially constrained, and the unconstrained indicator $1_{ (e,z) \in \mathscr{C}^c }$ is
\begin{align*}
    1_{(e, z) \in \mathscr{C} c}&=1\left\{\left(\frac{\left(r^{l}-r\right) z \alpha}{w}\right)^{\frac{1}{1-\alpha}}<\left(\frac{\phi e}{z}\right)^{1 / \alpha}\right\}.
\end{align*}
At the reflecting barriers, we have
$$\partial_{z} v(e, \underline{z})=\partial_{z} v(e, \bar{z})=0, \quad \text { for any } e.$$
We also have the boundary conditions
$$v(e, z) = \underline{v}(e) = 0.01, \quad \text{when } e = 0.01.$$
Banks will exit the market with zero equity, because $de_t \leq 0$ for $e_t=0$, and $\pi(0) = - c_f < 0 $, i.e., zero equity is an absorbing state, and continuing the operations when equity is zero will guarantee negative profit and thus is worse than exiting the market.

Finally, we describe entry dynamics. Banks first decide whether or not to enter, and then draw their productivity from the distribution $\psi(z)$. We assume that entry incurs a one-time cost $c_{e}$, and each entrant has the same initial equity of $e_0$. The mass of firms entering the market is determined by 
\[
m = \bar{m} \exp \left(\beta_M\left(\iint_{e,z}v(e,z)\psi(e,z)dzde - c_e\right)\right).
\]
The above is a softer version of the free-entry condition. When $\beta_M \to \infty $, entry incentive with respect to entry benefit is going to infinity, so the present value of entry must be zero and we arrive at the free-entry condition
\[\iint_{e,z}v(e,z)\psi(e,z)dzde-c_{e}=0.\]

\paragraph{Solve for invariant distribution and estimation.}
Denote the stationary bank distribution as $g(e,z)$. This distribution does not include banks that exit the market, so we have 
\[ g(e,z) =  1_{ v(e,z) > \underline{v}(e) } g(e,z), \]
where the assumption is that when banks are indifferent between staying or exiting the market, they choose to exit the market.

The Kolmogorov forward equation (KFE) for the stationary distribution in banking industrial dynamic model is
\begin{equation} \label{eq:kfe}
    0 = -\frac{\partial}{\partial z}\left(\mu_z(z)g(e,z)\right)-\frac{\partial}{\partial e }\left(\mu_{e}(e,z)g(e,z)\right)+\frac{1}{2}\frac{\partial^2}{\partial z^2}\left(\sigma(z)^2g(e,z)\right)+m\psi (e,z), \quad v(e,z)>\underline{v}(e).
\end{equation}
With the stationary distribution, we can get the aggregate loan
\[L=  \iint g(e,z) f(z, l^{*}(e,z)) dzde.\]
Furthermore, the equilibrium loan spread is determined by the household loan demand function
\begin{equation} \label{eq:rl-r}
    r^l - r  = \beta (L+D_0)^{-\varepsilon}
\end{equation}
for $D_0 = \iint_{e,z}d(e,z)g(e,z)dzde$ (here $g$ is not normalized), where $d = z(l^*)^{\alpha}$. We assume that $\psi(e,z)$ is a truncated normal distribution ($\Phi(\cdot)$ is the cumulative distribution function (CDF) of normal distribution)
\[
\psi(e,z) = \frac{1}{\bar{e}\times(\Phi(\frac{\overline{z}-z_{m}}{\sigma_{\psi}})-\Phi(\frac{\underline{z}-z_{m}}{\sigma_\psi}))}\frac{1}{\sqrt{2\pi \delta_\psi^2}}\exp\left(-\frac{(z-z_{m})^2}{2\delta_{\psi}^2}\right)\times 1_{e<\bar{e}}\times 1_{z\in[\underline{z},\overline{z}]}.
\]

\paragraph{Boundary conditions.}
(1) Banks exit at $\underline{v}(e) = e$ (absorbing boundary), which means $g(e,z) = 0$ when $v(e,z)= e$. (2) Reflecting boundary for stochastic productivity $z$: $-\mu_{z}(z)g(e,z)+\frac{1}{2}\frac{\partial }{\partial z} (\sigma(z)^2 g(e,z)) = 0$, when $z = \underline{z},\overline{z}$.

\paragraph{Specification.}
Model parameter specifications are shown in Table~\ref{tab:iofin_parameter}, and additional problem setup details can be found in Appendix~\ref{secApx:firm dynamic}.

\begin{table}[htbp]
\centering
\caption{\textbf{Parameter specification of the model in Section \ref{sec:firm dynamic setup}.}}
\label{tab:iofin_parameter}
\begin{tabular}{l|l}
\toprule Description & Value \\
\midrule
Bank equity payout rate & $\kappa = 0.005$ \\
Share of labor & $\alpha=0.3$ \\
Leverage constraint parameter & $\phi=10$\\
Benchmark interest rate & $r = 0.03$ \\
Fixed operating cost & $c_f = 0.03$\\
Entry cost & $c_e = 0.1$\\
Boundary and mean of productivity & $\underline{z} = 0.2$, $\overline{z} = 10$, $z_{m}=5$\\
Lower and upper bound of state space & $e_{\text{min}} = 0.01$, $e_{\text{max}} = 1.2$\\
Drift and volatility of $z$ & $\mu(z) = -0.005(z-z_m)$, $\sigma(z) = 0.08$\\
Deposit/Loan supply function's constant & $D_0 = L_0 = 1.0$\\
Entrance distribution  parameters&  $\sigma_{\psi}=\frac{\overline{z}-\underline{z}}{4}$, $\bar{e}=0.15$, $\bar{m}=0.1$ \\
Entrance elasticity & $\beta_M = 1\times 10^3$\\
\bottomrule
\end{tabular}
\end{table}

\subsection{Solving the model}
\label{sec:hjb-forward}

Our goal is to solve for $v(e, z)$ and $g(e, z)$. In Section~\ref{sec:elim-m}, we first eliminate the role of $m$ from the model. Then we present some technical details and the results in the following sections.

\subsubsection{Elimination of $m$}
\label{sec:elim-m}

The only equation in which $m$ directly shows up is the KFE in Eq.~\eqref{eq:kfe}, in which $g(e, z)$ scales linearly in $m$. Linear scaling of $g(e, z)$ does not affect the values of the moment targets, because $g(e, z)$ is normalized in those calculations. However, scaling $g(e, z)$ affects Eq.~\eqref{eq:rl-r}, in which
\[L = \iint_{e, z} g(e, z) f(z, l^*(e, z))dzde\] is scaled by the same factor. Our goal is to exactly satisfy Eq.~(\ref{eq:rl-r}) by scaling $g(e, z)$. Afterwards, in order to still satisfy the KFE, we scale $m$ by the same factor. Using the scaling property to solve $m$ is a standard approach in firm dynamics literature, i.e., \citep{hopenhayn1992entry}.

To implement this idea with a PINN, we fix $m = 1$ throughout training.  After training, let $\mathcal{N}_g(e, z)$ be the PINN-predicted value of $g(e, z)$. We calculate 
\[\mathcal{N}_L = \iint_{e, z} \mathcal{N}_g (e, z) f(z, l^*(e, z))dzde.\]
To automatically satisfy Eq.~(\ref{eq:rl-r}), we set
$$m = \frac{\frac{\beta}{r^l-r}-D_0}{L}$$
due to 
$$r^l - r = \beta(\mathcal{N}_Lm + D_0)^{-\varepsilon},$$
in which we use $\varepsilon = 1$. After solving for $m$, our final prediction of $g(e, z)$ is $g(e, z) = m\mathcal{N}_g(e, z)$.

\subsubsection{Other technical details}
\label{sec:hjb-details}

We solve the model with the unknown endogenous boundary $r^l$. When training the neural network, we enforce the Dirichlet boundary condition on $v$ via a soft boundary condition and the Dirichlet boundary condition on $g$ through a hard boundary condition. Furthermore, we enforce the Neumann boundary condition on $g$ via a soft boundary condition. We use loss weights of $10^6$ for the HJB residual, $5 \times 10^4$ for the KFE residual, $10^3$ for the free-entry condition, $10^2$ for the Dirichlet boundary condition on $v$, $10^3$ for the Neumann boundary condition on $v$, and $10^5$ for the Neumann boundary condition on $g$. 

Additionally, we train with $2^{16}$ training points sampled inside the domain, $2^{10}$ training points sampled on the boundary, and $2^{16}$ points sampled inside the domain for testing. When estimating $r^l$, we scale it up $100$ times while training and scale it back down afterwards. 

\subsubsection{Results}

\begin{figure}[htbp]
    \centering
    \includegraphics[width=\textwidth]{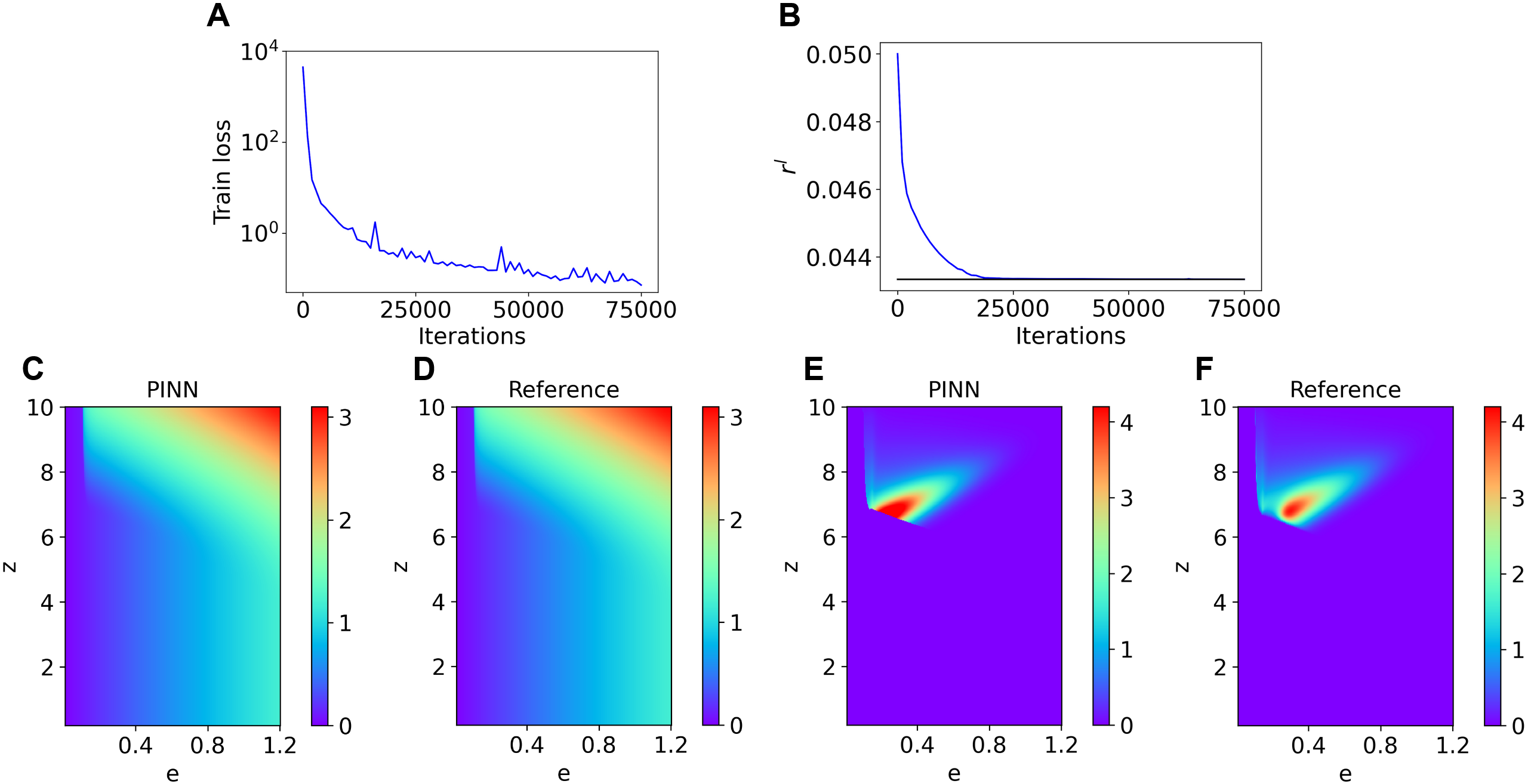}
    \caption{\textbf{PINN results for the forward HJB equation.} (\textbf{A}) The value of training loss throughout training. (\textbf{B}) The convergence of $r^l$ throughout training. (\textbf{C}) Predicted solution for $v$ using PINN. (\textbf{D}) Reference solution for $v$ using the finite difference method. (\textbf{E}) Predicted solution for $g$ using PINN. (\textbf{F}) Reference solution for $g$ using the finite difference method.}
    \label{fig:hjb-forward}
\end{figure}

The training results are displayed in Fig.~\ref{fig:hjb-forward}. The training loss decreases steadily (Fig.~\ref{fig:hjb-forward}A), and the endogenous variable $r^l$ quickly converges to its true value within around 20000 iterations (Fig.~\ref{fig:hjb-forward}B). As shown in Figs.~\ref{fig:hjb-forward}C and \ref{fig:hjb-forward}D, the PINN prediction and reference solution for $v(e, z)$ are very similar visually. In fact, the $L^2$ relative error for $v(e, z)$ is $0.54\%$. In Figs.~\ref{fig:hjb-forward}E and \ref{fig:hjb-forward}F, we show the PINN prediction and reference solution have a similar shape, though the PINN prediction appears shifted to the left. Here, we compute the $L^2$ relative error of $g_c$, which is defined as the CDF of $g$, i.e., $g_c(e,z) = \int_{\underline{z}}^{z}\int_{0}^eg(e',z')de'dz'$, to be $4.32\%$. The final predicted values of $r^l$ and $m$ after training are listed in Table~\ref{tab:hjb-forward}, and both have very good accuracy (0.01\% and 2.12\%, respeticely). 

\begin{table}[htbp]
    \centering
    \caption{\textbf{Predicted values of endogenous variables $r^l$ and $m$ after training.}}
    \label{tab:hjb-forward}
    \begin{tabular}{c|ccc}
        \toprule
         Variable & True value & Predicted value & Relative error \\
         \midrule
         $r^l$ & 0.043337 & 0.043343 & 0.01\% \\
         $m$ & 0.2194 & 0.2242 & 2.12\% \\
         \bottomrule
    \end{tabular}
\end{table}

\subsection{Simultaneously solving and estimating the model}
\label{sec:firm dynamic-estimation}

In practice, we may want to estimate some unknown parameters of the model, which requires additional information in the form of moment targets. Here, we aim to predict the value of $c_e$, along with several other parameters. We note that $c_e$ only appears in the free-entry condition. To take advantage of this, we train the PINN without the free-entry condition, and after training, we predict $c_e$ by calculating 
\[c_e = \iint_{e, z} v(e, z) \psi(e, z) dzde.\]

\subsubsection{Estimation of two parameters}
\label{sec:hjb-inv-two-var}

In this example, we estimate the two parameters $\alpha$ and $c_e$. Define the density $g'(e,z)$ as $\max\{g(e,z),0\}$ discounted by normalization factor: $\frac{1}{\iint _{e,z}\max\{g(e,z),0\}dzde}$, representing the normalized density. We note that as the KFE operator $\hat{L}^*$ is the Markov process's generator, the distribution is always positive when evolving over time; however, in numerical exercise, $g(e,z)$ at some point can be negative. In estimation, we train the unknown parameters to match the average productivity $z_{\text{target}}$ and average labor $l^*_{\text{target}}$:
\begin{align*}
z_{\text{target}}&=\iint_{e,z}zg'(e,z)dzde, \\
l^*_{\text{target}} &= \iint_{e,z}l^* g'(e,z)dzde.
\end{align*}

In our implementation, we enforce the Dirichlet boundary condition on $v$ via a soft boundary condition and the Dirichlet boundary condition on $g$ through constructing the surrogate solution of $g$ as
\[g(e, z) = \mathcal{N}_g(e, z) 1_{v(e, z) > e},\]
where $\mathcal{N}_g(e, z)$ is a neural network. Furthermore, we enforce the Neumann boundary condition on $g$ via a soft boundary condition. Specifically, we use loss weights of $10^6$ for the HJB residual, $5 \times 10^4$ for the KFE residual, $10^1$ for $z_{\text{target}}$, $10^6$ for $l^*_{\text{target}}$, $10^2$ for the Dirichlet boundary condition on $v$, $10^3$ for the Neumann boundary condition on $v$, and $10^5$ for the Neumann boundary condition on $g$. Furthermore, we train with $2^{14}$ points sampled inside the domain and $2^9$ points sampled on the boundary. When estimating $\alpha$, we scale it up $10$ times while training and scale it back down afterwards. 

\begin{figure}[htbp]
    \centering
    \includegraphics[width=\textwidth]{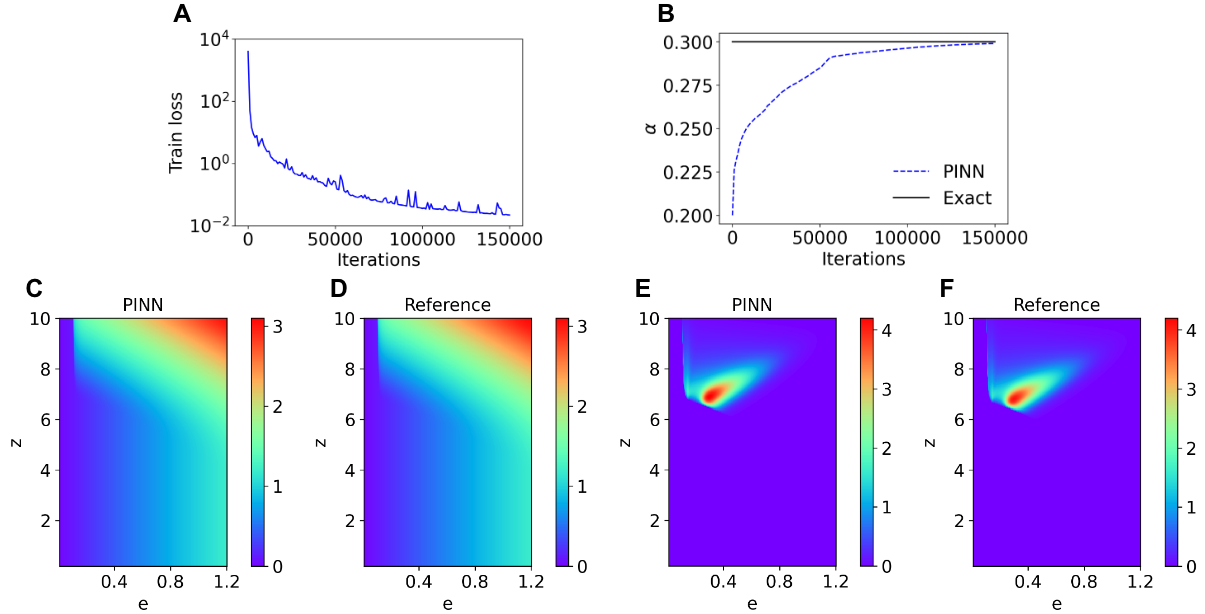}
    \caption{\textbf{PINN results for the inverse HJB equation with estimation of two parameters.} (\textbf{A}) The value of train loss throughout training. (\textbf{B}) The convergence of $\alpha$ throughout training. (\textbf{C}) Predicted solution for $v$ using PINN. (\textbf{D}) Reference solution for $v$. (\textbf{E}) Predicted solution for $g$ using PINN. (\textbf{F}) Reference solution for $g$.}
    \label{fig:hjb-inv-two-var}
\end{figure}

The training results are displayed in Fig.~\ref{fig:hjb-inv-two-var}. In addition, the predictions and errors for $c_e$ and $\alpha$ are displayed in Table~\ref{tab:hjb-inv-two-var}, along with the predictions and errors for $z_{\text{target}}$ and $l^*_{\text{target}}$. The training loss decreases steadily (Fig.~\ref{fig:hjb-inv-two-var}A), and the $L^2$ relative error of $v$ ends at $3.28\%$. The trajectory of $\alpha$ is displayed in Fig.~\ref{fig:hjb-inv-two-var}B. As shown in Figs.~\ref{fig:hjb-inv-two-var}C and D, the PINN predicts the solution for $v$ accurately. Furthermore, Figs.~\ref{fig:hjb-inv-two-var}E and F demonstrate that the PINN predicts the solution for $g$ accurately. The errors of both the predicted parameters and the moments of the predicted solution are under $5\%$ (Table~\ref{tab:hjb-inv-two-var}). Finally, the $L^2$ relative errors of $v$ and $g_c$ are $3.28\%$ and $4.24\%$, respectively.

\begin{table}[htbp]
    \centering
    \caption{\textbf{PINN results for the inverse HJB equation with estimation of two parameters.} Both unknown parameters ($c_e$ and $\alpha$) and moment targets ($z_\text{target}$ and $l^*_\text{target}$) are shown.}
    \label{tab:hjb-inv-two-var}
    \begin{tabular}{c|ccc}
    \toprule
     & True value & Predicted value & Relative error \\
    \midrule
    $c_e$ & 0.1 & 0.0959 & 4.10\% \\
    $\alpha$ & 0.3 & 0.299 & 0.31\% \\
    $z_\mathrm{target}$ & 7.524 & 7.611 & 1.15\% \\
    $l^*_\mathrm{target}$ & 0.00647 & 0.00637 & 1.46\% \\
    \bottomrule
    \end{tabular}
\end{table}

\subsubsection{Estimation of three parameters}
\label{sec:hjb-inv-three-var}

Next, we estimate three parameters: $\alpha, c_e,$ and $c_f$. We will match four moment conditions: average equity $e_{\text{target}}$, average productivity $z_{\text{target}}$, average leverage $\ell_{\text{target}}$, and average labor $l^*_{\text{target}}$, defined as
\begin{align*}
    e_{\text{target}}&=\iint_{e,z}eg'(e,z)dzde, \\
    z_{\text{target}}&=\iint_{e,z}zg'(e,z)dzde, \\
    \ell_{\text{target}}&=\iint_{e,z}\frac{f(z,l^*(e,z))}{e}g'(e,z)dzde, \\
    l^*_{\text{target}} &= \iint_{e,z}l^* g'(e,z)dzde.
\end{align*}

In our implementation, we mostly use the same setup as Section~\ref{sec:hjb-inv-two-var}. In addition to the previous loss weights, we use loss weights of $10^2$ for $e_\text{target}$ and 1 for $\ell_\mathrm{target}$. We also use more training points: $2^{16}$ training points sampled inside the domain and $2^{10}$ training points sampled on the boundary. Additionally, we scale $c_f$ up by $100$ times, and we scale it back down after training.

\begin{figure}[htbp]
    \centering
    \includegraphics[width=16.5cm]{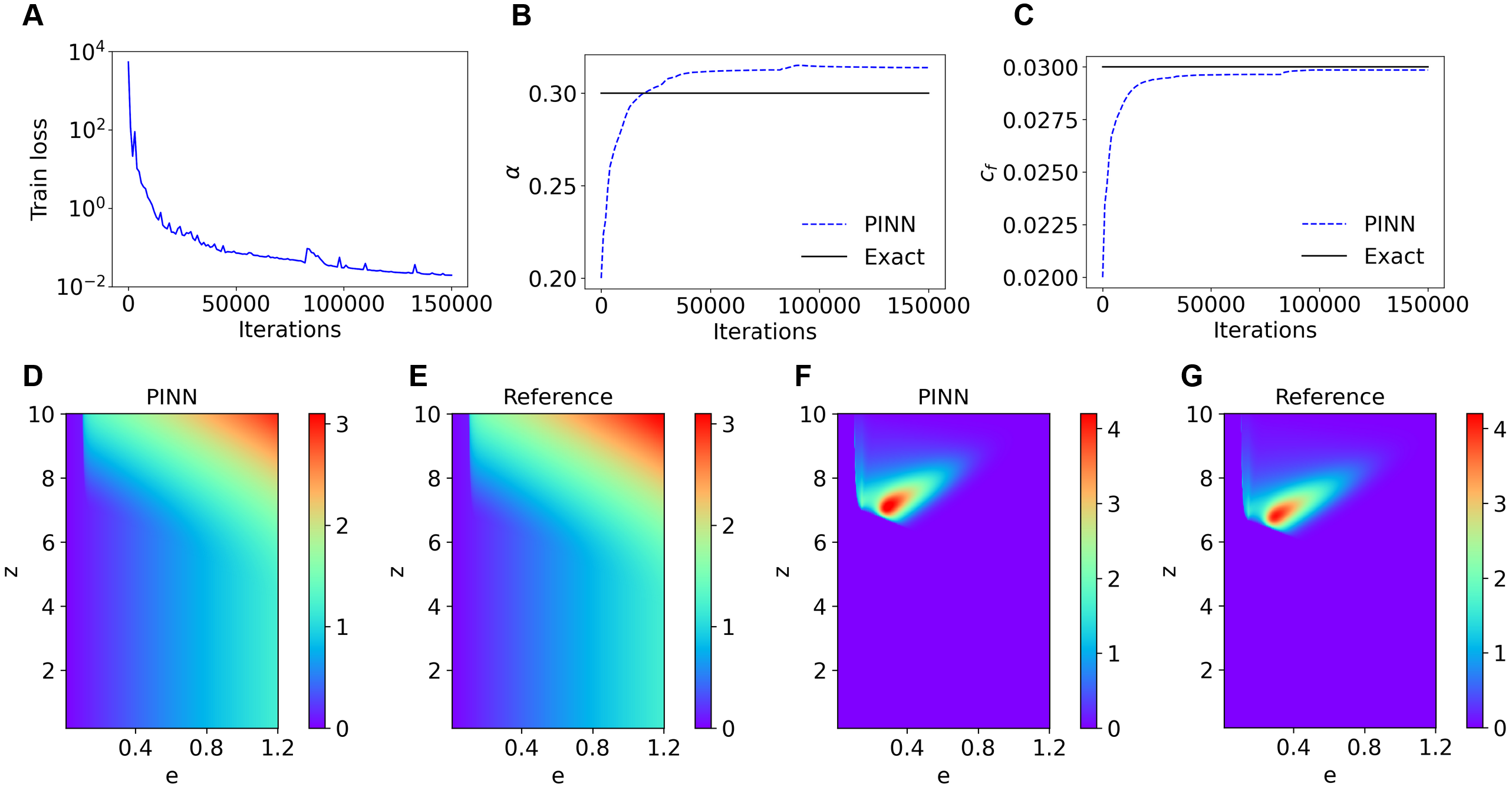}
    \caption{\textbf{PINN results for the inverse HJB equation with estimation of three parameters.} (\textbf{A}) The value of train loss throughout training. (\textbf{B}) The convergence of $\alpha$ throughout training. (\textbf{C}) The convergence of $c_f$ throughout training. (\textbf{D}) Predicted solution for $v$ using PINN. (\textbf{E}) Reference solution for $v$. (\textbf{F}) Predicted solution for $g$ using PINN. (\textbf{G}) Reference solution for $g$.}
    \label{fig:hjb-inv-three-var}
\end{figure}

The training results are displayed in Fig.~\ref{fig:hjb-inv-three-var}. In addition, the predictions and errors for $c_e$, $\alpha$, and $c_f$ are displayed in Table~\ref{tab:hjb-inv-three-var}, along with the predictions and errors for $e_{\text{target}}$, $z_{\text{target}}$, $\ell_{\text{target}}$, and $l^*_{\text{target}}$. The training loss decreases steadily (Fig.~\ref{fig:hjb-inv-three-var}A). We see that both $\alpha$ and $c_f$ converge to a values close to the true values (Figs.~\ref{fig:hjb-inv-three-var}B and C). As shown in Figs.~\ref{fig:hjb-inv-three-var}D and E, the PINN predicts the solution for $v$ accurately. Furthermore, Figs.~\ref{fig:hjb-inv-three-var}F and G demonstrate that the PINN predicts the solution for $g$ accurately. The errors of both the predicted parameters and the moments of the predicted solution are under $10\%$ (Table~\ref{tab:hjb-inv-three-var}). Finally, the $L^2$ relative errors of $v$ and $g_c$ are $8.53\%$ and $13.21\%$, respectively.

\begin{table}[htbp]
    \centering
    \caption{\textbf{PINN results for the inverse HJB equation with estimation of three parameters.} Unknown parameters $c_e$, $\alpha$, and $c_f$ as well as moment targets $e_\text{target}$, $z_\text{target}$, $\ell_\text{target}$ and $l^*_\text{target}$ are shown.}
    \label{tab:hjb-inv-three-var}
    \begin{tabular}{c|ccc}
    \toprule
     & True value & Predicted value & Relative error \\
    \midrule
    $c_e$ & 0.1 & 0.0934 & 6.60\% \\
    $\alpha$ & 0.3 & 0.314 & 4.59\% \\
    $c_f$ & 0.03 & 0.0299 & 0.49\% \\
    $e_\mathrm{target}$ & 0.418 & 0.385 & 7.96\% \\
    $z_\mathrm{target}$ & 7.524 & 7.786 & 3.47\% \\
    $\ell_\mathrm{target}$ & 4.697 & 4.955 & 5.50\% \\
    $l^*_\mathrm{target}$ & 0.00647 & 0.00677 & 4.70\% \\
    \bottomrule
    \end{tabular}
\end{table}
\section{A macroeconomic model with the financial sector}
\label{sec:brunnermeier sannikov}

The next problem we discuss describes the model in \cite{brunnermeier2014macroeconomic}. The problem features occasional binding constraints, non-linear financial amplifications, and singularity at the boundary. Here, we describe the model only briefly, see \cite{brunnermeier2014macroeconomic} for the detailed model setup.

\subsection{Problem setup}
\label{sec:brunnermeier sannikov setup}

There are two types of agents, a continuum mass of bankers and a continuum mass of households, both with risk-neutral utility. There are two types of assets: productive capital and risk-free asset. Per unit of capital, banker productivity is $a$ while household productivity is $\underline{a} < a$.   Both bankers and households can borrow and lend at the risk-free rate. Banker discount rate $\delta$ is smaller than the household discount rate, which motivates them to borrow from households.

To solve for the equilibrium, we need to solve the value of capital  $q(\omega)$ and the endogenous marginal value of wealth $\theta(\omega)$ for the banker, where $\omega$ is the fraction of the banker wealth among total wealth and is the state variable that determines asset prices and allocations. There is an endogenous boundary $\eta^{*}$ where banks pay dividends. The function $q(\eta)$ is increasing and $\theta(\eta)$ is decreasing over $[0,\eta^{*}]$, with boundary conditions
\[
q(0)=\underline{q},\quad\theta(\eta^{*})=1,\quad q'(\eta^{*})=0,\quad\theta'(\eta^{*})=0,\quad\text{and \ensuremath{\quad\lim_{\eta\to0}\theta(\eta)=\infty}},
\]
where the boundary value $\underline{q}$ is determined by 
\begin{equation}
\underline{q}=\max_{q}\frac{\underline{a}-\iota(x)}{r-(\Phi(x)-\underline{\delta})}. \label{eq:q_}
\end{equation}
Note that the maximum possible value of $q(\eta)$ is determined by the following equation:
\[ \frac{{a}-\iota(\bar{q})}{\bar{q}} + \Phi(\bar{q}) - {\delta} - r =0,  \]
where $\iota(\cdot)$ and $\Phi(\cdot)$ are given by
\[
\Phi(x)=\frac{x-1}{\kappa} \quad \text{and} \quad \iota(x)=\Phi(x)+\frac{1}{2}\kappa\Phi(x)^{2}.
\]
Functions $q(\eta)$ and $\theta(\eta)$ are second-order ODEs. Thus, we need two boundary conditions for $q(\eta)$ and two boundary conditions for $\theta(\eta)$. The above are five conditions, but we also have an unknown boundary $\eta^{*}$. 
To solve the model, we first find $\psi\in(\eta,\eta+q(\eta)/q'(\eta))$, such that
\begin{equation} \label{eq:psi}
\frac{a-\underline{a}}{q(\eta)}+\underline{\delta}-\delta+(\sigma+\sigma^{q}(\eta))\sigma^{\theta}(\eta)=0,
\end{equation}
where 
\[
\sigma^{\eta}(\eta)\eta=\frac{(\psi-\eta)\sigma}{1-(\psi-\eta)q'(\eta)/q(\eta)}, \quad \sigma^{q}(\eta)=\frac{q'(\eta)}{q(\eta)}\sigma^{\eta}(\eta)\eta, \quad \text{and} \quad \sigma^{\theta}(\eta)=\frac{\theta'(\eta)}{\theta(\eta)}\sigma^{\eta}(\eta)\eta.
\]
We can easily prove that the left side of Eq.~\eqref{eq:psi} is monotonic in $\psi$ and has a unique solution. If the above solution indicates $\psi>1$, then we set $\psi=1$ and recalculate $\sigma^{\eta}(\eta)$, $\sigma^{q}(\eta)$, and $\sigma^{\theta}(\eta)$. 

Next, we compute the second-order derivatives, via 
\begin{equation}
    q''(\eta)=\frac{2\left(\mu^{q}(\eta)q(\eta)-q'(\eta)\mu^{\eta}(\eta)\eta\right)}{\sigma^{\eta}(\eta)^{2}\eta^{2}},
    \label{eq:qpp}
\end{equation}
\begin{equation}
    \theta''(\eta)=\frac{2\left(\mu^{\theta}(\eta)\theta(\eta)-\theta'(\eta)\mu^{\eta}(\eta)\eta\right)}{\sigma^{\eta}(\eta)^{2}\eta^{2}},
    \label{eq:thetapp}
\end{equation}
where
\[
\mu^{\eta}(\eta)=-\frac{(\psi - \eta)(\sigma + \sigma^q(\eta))(\sigma+\sigma^{q}(\eta)+\sigma^{\theta}(\eta))}{\eta}+\frac{a-\iota(q(\eta))}{q(\eta)}+(1-\psi)(\underline{\delta}-\delta),
\]
\[
\mu^{q}(\eta)=r-\frac{a-\iota(q(\eta))}{q(\eta)}-\Phi(q(\eta))+\delta-\sigma\sigma^{q}(\eta)-\sigma^{\theta}(\eta)\cdot(\sigma+\sigma^{q}(\eta)), \quad \text{and} \quad \mu^{\theta}=\rho-r.
\]

Model parameter specifications are shown in Table~\ref{tab:macrofin_parameter} and additional problem setup details and derivations can be found in Appendix~\ref{secApx:MacroFin}.




\begin{table}[htbp]
\centering
\caption{\textbf{Parameter specification of model in Section \ref{sec:brunnermeier sannikov setup}.}}
\label{tab:macrofin_parameter}
\begin{tabular}{l|l}
\toprule Description & Value\\
\midrule Capital productivity (experts) & $a = 0.11$ \\
Capital productivity (households)  & $\underline{a} = 0.05$ \\
Discount rate (experts) & $\rho = 0.06$\\
Discount rate (households) & $r = 0.06$\\
Unit volatility of capital & $\sigma = 0.025$\\
Capital depreciation rate (experts) & $\delta = 0.03$\\
Capital depreciation rate (households) & $\underline{\delta}=0.08$\\
Loss factor in investment & $\kappa = 10$\\
\bottomrule
\end{tabular}
\end{table}

\subsection{Solving the model}
\label{sec:intermediary-solution}

Our goal is to solve for the functions $q$, $\theta$, and $\psi$. To make network training easier, we first introduce a few useful techniques below.

\subsubsection{Change of variable to deal with singularity}

First, we perform a change of variable for $\theta$ by defining the function $\hat{\theta}(\eta) = \frac{1}{\theta(\eta)}$. The motivation behind this is to deal with the singularity for $\theta$ at $\eta = 0$. With this change of variable, we rewrite all the boundary conditions:
\begin{align*}
    \lim_{\eta \to 0} \theta(\eta) = \infty, & \quad\text{i.e.,}\quad \hat{\theta}(0) = 0, \\
    \theta(\eta^*) = 1, & \quad\text{i.e.,}\quad \hat{\theta}(\eta^*) = 1, \\
    \theta'(\eta^*) = 0, & \quad\text{i.e.,}\quad \hat{\theta}'(\eta^*) = 0.
\end{align*}
We now rewrite the ODEs in the problem setup with the new variable $\hat{\theta}$. Because 
\[\sigma^\theta(\eta)\hat{\theta}(\eta) = -\hat{\theta}(\eta)'\sigma^\eta(\eta)\eta,\]
we rewrite Eq.~\eqref{eq:psi} as
\[\left(\frac{a-\underline{a}}{q} + \underline{\delta} - \delta\right) \hat{\theta}(\eta) + (\sigma + \sigma_q) \sigma^\theta(\eta) \hat{\theta}(\eta) = 0.\]
We also rewrite these definitions as
\begin{align*}
\mu^\eta(\eta)\eta\hat{\theta}(\eta) &= -(\psi - \eta) (\sigma + \sigma^q) (\hat{\theta}(\eta) (\sigma + \sigma^q) + \sigma^\theta(\eta)\hat{\theta}(\eta)) + \eta \hat{\theta}(\eta) \left(\frac{a - \iota}{q} + (1 - \psi) (\underline{\delta} - \delta)\right), \\
\mu^q(\eta)\hat{\theta}(\eta) &= \hat{\theta}(\eta) \left(r - \frac{a - \iota(q(\eta))}{q} - \Phi(q(\eta)) + \delta - \sigma \sigma^q \right) - \sigma^\theta(\eta)\hat{\theta}(\eta) \cdot (\sigma + \sigma^q(\eta)).
\end{align*}
Lastly, Eqs.~\eqref{eq:qpp} and \eqref{eq:thetapp} are now rewritten as
\begin{align*}
    &q''(\eta) \sigma^\eta(\eta)^2\eta^2 \hat{\theta}(\eta) = 2 (\mu^q(\eta)\hat{\theta}(\eta) q(\eta) - q'(\eta) \mu^\eta(\eta)\eta\hat{\theta}(\eta)), \\
    &\sigma^\eta(\eta)^2\eta^2 \cdot (2  \hat{\theta}'(\eta)^2 - \hat{\theta}(\eta) \hat{\theta}''(\eta)) = 2(\mu^\theta \hat{\theta}(\eta)^2 + \hat{\theta}'(\eta) \mu^\eta(\eta)\hat{\theta}(\eta)).
\end{align*}

\subsubsection{Explicitly solving for $\psi$}

In this problem, we simultaneously solve for $q$ and $\hat{\theta}$. As $\psi$ can be solved analytically given $q$ and $\hat{\theta}$, and thus we update $\psi$ every 1000 iterations given the current $q$ and $\hat{\theta}$. To compute $\psi$ during training, we solve for the function explicitly. Specifically, as
$$\sigma^{q}(\eta) =\frac{q'(\eta)}{q(\eta)}\sigma^{\eta}(\eta)\eta = \frac{q'(\eta)(\psi - \eta)\sigma}{q(\eta) - (\psi - \eta)q'(\eta)},$$
we have
$$\sigma^q(\eta) + \sigma 
= \frac{q'(\eta)(\psi - \eta)\sigma}{q(\eta) - (\psi - \eta)q'(\eta)} + \sigma
= \frac{q'(\eta)(\psi-\eta)\sigma + \sigma q(\eta) - \sigma(\psi - \eta)q'(\eta)}{q(\eta) - (\psi - \eta)q'(\eta)} 
= \frac{\sigma q(\eta)}{q(\eta) - (\psi - \eta)q'(\eta)},$$
and
$$(\sigma + \sigma^q(\eta))\sigma^{\theta}(\eta)
= \left(\frac{\sigma q(\eta)}{q(\eta) - (\psi - \eta)q'(\eta)}\right)\frac{\theta'(\eta)(\psi - \eta)\sigma}{\theta(\eta)(1-(\psi-\eta)q'(\eta)/q(\eta))}
= \frac{\theta'(\eta)}{\theta(\eta)}\left(\frac{(\sigma q(\eta))^2(\psi - \eta)}{(q(\eta)-(\psi-\eta)q'(\eta))^2}\right).$$
Using Eq.~\eqref{eq:psi}, we have
$$\theta(\eta)(q(\eta)-(\psi-\eta)q'(\eta))^2(a-\underline{a} + q(\eta)(\underline{\delta} - \delta)) + \sigma^2q(\eta)^3(\psi - \eta)\theta'(\eta) = 0,$$
from which we solve for $\psi$ using the quadratic formula.

\subsubsection{Other technical details}
\label{sec:brunnermeier-sannikov-tech-detail}

We employ several additional techniques when constructing the model to improve its results. To satisfy the boundary conditions, we enforce hard boundary conditions by constructing the solution as
\begin{align*}
    q(\eta) &= (\eta - \eta^*)^2\mathcal{N}_q(\eta) + C, \\
    \hat{\theta}(\eta) &= \eta(\eta - \eta^*)^2\mathcal{N}_{\hat{\theta}}(\eta) - (\eta/\eta^*)^2 + 2\eta/\eta^*,
\end{align*}
where $C$ is a trainable variable that represents $q(\eta^*)$, while $\mathcal{N}_q(\eta)$ and $\mathcal{N}_{\hat{\theta}}(\eta)$ are the first and second components of the network output, respectively. This automatically satisfies the boundary conditions $q'(\eta^*) = 0$, $\hat{\theta}(0) = 0$, $\hat{\theta}(\eta^*) = 1$, and $\hat{\theta}'(\eta^*) = 0$. 
For the boundary condition $q(0) = \underline{q}$, we use a soft boundary condition. We train two variables $C = q(\eta^*)$ and $\eta^*$, and we set initial guesses of $C = 1.0$ and $\eta^* = 0.4$. While the value of $C$ is flexible, the initial guess for $\eta^*$ comes from experimentation.

As $q(\eta)$ features a steep gradient for small $\eta$, to capture this gradient better, we use a multiscale feature layer \citep{wang2020multi,yazdani2020systems} of $(\eta, 2\eta, 3\eta, \dots, 10\eta)$, which makes it easier to capture the sharp increase for small $\eta$. We also add additional ten points in the interval $[0, 10^{-4}]$.

We want to ensure that $q$ is increasing and $\theta$ is decreasing. By the change of variable, $\theta$ decreasing is equivalent to $\hat{\theta}$ increasing, so to enforce $q$ increasing and $\hat{\theta}$ increasing, we add loss terms
\begin{align*}
    \mathcal{L}_q = \int_0^{\eta^*} (\min(q'(\eta), 0))^2 d\eta \quad \text{and} \quad \mathcal{L}_{\hat{\theta}} = \int_0^{\eta^*} (\min(\hat{\theta}'(\eta), 0))^2 d\eta,
\end{align*}
representing loss from the mean squared value of $\min(q'(\eta), 0)$ and $\min(\hat{\theta}'(\eta), 0).$ This pushes the functions to have positive derivatives, so they are increasing. We note that for $\underline{q}$ in Eq.~(\ref{eq:q_}), it is easy to find that $\underline{q} = 0.4862$.

We use loss weights of $10^6$ for the ODE system, $q$ increasing, and $\hat{\theta}$ increasing losses, and a loss weight of $1$ for the boundary condition $q(0) = \underline{q}$. Furthermore, we train with $10^3$ points sampled inside the domain and $2$ points sampled on the boundary. While training, we use two different learning rate schedulers. The first learning rate scheduler has an initial learning rate of $1 \times 10^{-3}$ with a decay rate of $0.5$ and decay step of $1500$. We use this scheduler for the first $1 \times 10^5$ iterations. Then, we switch to a learning rate scheduler with initial learning rate $1 \times 10^{-5}$, decay rate $0.5$, and decay step $1500$ for another $2 \times 10^5$ iterations.

\subsubsection{Results}

\begin{figure}[htbp]
    \centering
    \includegraphics[width=\textwidth]{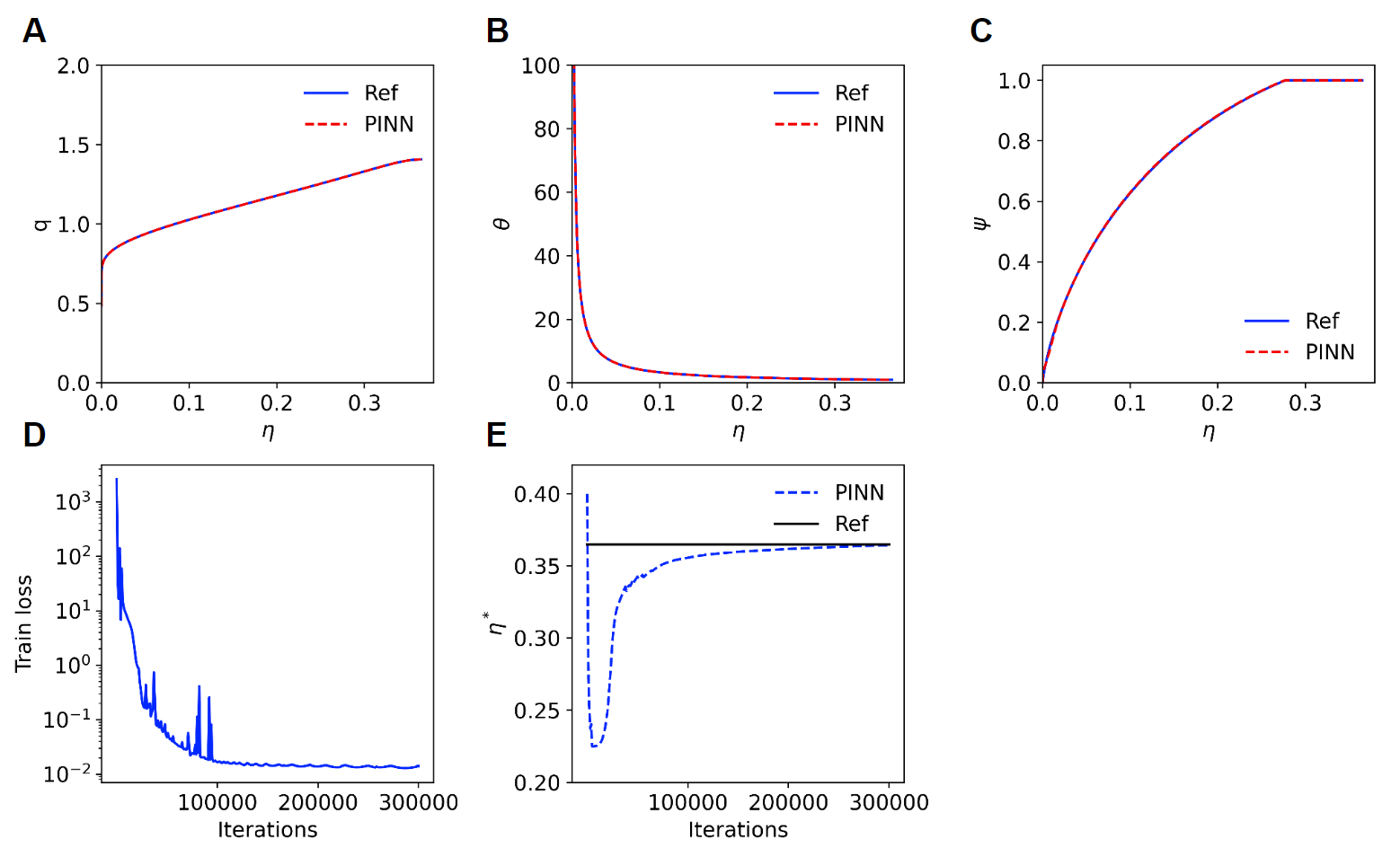}
    \caption{\textbf{PINN results for the forward B\&S model.} (\textbf{A}) The reference value and prediction for $q$ after training. (\textbf{B}) The reference value and prediction for $\theta$ after training. (\textbf{C}) The reference value and prediction for $\psi$ after training. (\textbf{D}) The value of train loss throughout training. (\textbf{E}) The convergence of $\eta^*$ throughout training.}
    \label{fig:iap_forward} 
\end{figure}

The training results are displayed in Fig.~\ref{fig:iap_forward}, while the predicted value of $\eta^*$ is displayed in Table~\ref{tab:bs_forward_pred}. In Fig.~\ref{fig:iap_forward}D, we see the training loss steadily decrease throughout training, suggesting the PINN converges to a solution. In Figs.~\ref{fig:iap_forward}A, B, and C, we see that visually, the reference and PINN predictions for $q$, $\theta$, and $\psi$ are nearly identical. The prediction for $q$ (Fig.~\ref{fig:iap_forward}A) has an $L^2$ relative error of $1.14\%$. The prediction for $\theta$ (Fig.~\ref{fig:iap_forward}B) has an $L^2$ relative error of $0.258\%$. The prediction for $\psi$ (Fig.~\ref{fig:iap_forward}C) has an $L^2$ relative error of $0.273\%$. We see the endogenous variable $\eta^*$ quickly converge to the true value (Fig.~\ref{fig:iap_forward}E). This is confirmed by the true and predicted values seen in Table \ref{tab:bs_forward_pred}, with an error of approximately $0.1\%$.

\begin{table}[htbp]
    \centering
    \caption{\textbf{Predicted value of $\eta^*$ after training.}}
    \label{tab:bs_forward_pred}
    \begin{tabular}{c|cc}
        \toprule
         Variable & True & Prediction \\
         \midrule
         $\eta^*$ & 0.3648 & 0.3644 \\
         \bottomrule
    \end{tabular}
\end{table}

\subsection{Simultaneously solving and estimating the model} \label{sec:intermediary-estimation}

In this section, we simultaneously solve the model and also estimate one or two parameters of the model. 

\subsubsection{Estimation of one parameter}
\label{sec:iap-one-param}

First, we estimate only one parameter. Specifically, we assume $a$ is unknown and impose a moment condition. There are two ways to match the moment condition. The first method is through simulation, i.e., we simply simulate $\eta_t$ according to its law of motion, 
\[ \frac{d\eta_{t}}{\eta_{t}}=\mu^\eta(\eta_{t})dt+\sigma^{\eta}(\eta_{t})dB_{t}, \]  
and then take the average moment value to match the target. A more method is through the Kolmogorov forward equation (KFE), which is more efficient and used in this study. We can solve for the KFE that describes the ODE of the stationary density function of $\eta_t$, $f(\eta)$: 
\[
-\frac{\partial}{\partial\eta}\left(\mu^\eta(\eta)\eta f(\eta)\right)+\frac{\partial^{2}}{\partial\eta^{2}}\left(\frac{1}{2}\sigma^{\eta}(\eta)^{2}\eta^2f(\eta)\right)\equiv\frac{\partial}{\partial \eta} J(\eta)=0,
\] 
with boundary condition $-\mu^\eta\eta f(\eta)+\frac{1}{2}\frac{\partial}{\partial \eta}\left(\sigma^\eta(\eta)^2f(\eta)\right)|_{\eta = \eta^*} = 0$. For simplicity, we define the function $Q$ as
\[
Q(\eta) = \frac{2\mu^\eta(\eta)\eta - \frac{\partial }{\partial \eta}(\sigma^\eta(\eta)^2\eta^2)}{\sigma^\eta(\eta)^2\eta^2}.
\]
Then, according to Appendix~\ref{secApx:Dimension Reduction Trick}, the density function can be written as
\[
f(\eta)=A\exp\left(-\int_{\eta}^{\eta^*} Q(\eta')d\eta'\right),
\]
where $A$ is the normalization factor
\[
\frac{1}{A}= \int_0^{\eta^*}\exp\left(-\int_{\eta'}^{\eta^*} Q(\eta)d\eta\right)d\eta'.
\]
We match the moment target
\begin{align*}
    a_{\text {target}} = \int_{0}^{\eta^{*}}[\psi(\eta) a+(1-\psi(\eta)) \underline{a}] f(\eta) d \eta
\end{align*}
where we have $a_{\text{target}} = 0.1095$. To compute the integrals in the moment conditions, we use numerical integration.

For the inverse problem, we use the same techniques used in the forward problem. Moreover, during training, we scale $a$ up by $10$, and we scale it back down after training, and we restrict $\eta^*$ to $(0.3, 0.5)$. Also, we use loss weights as $5\times 10^4$ for the ODE on $q''$, $10^4$ for the ODE on $\theta''$, $10^5$ for $q$ increasing, $10^3$ for $\theta$ decreasing, $10^4$ for the $a_\text{target}$ moment condition, and $10^0$ for the $q(0)=\underline{q}$ boundary condition. We train with $2^{13}-2$ points in the domain and $2$ points on the boundary. 

\begin{figure}[htbp]
    \centering
    \includegraphics[scale=0.28]{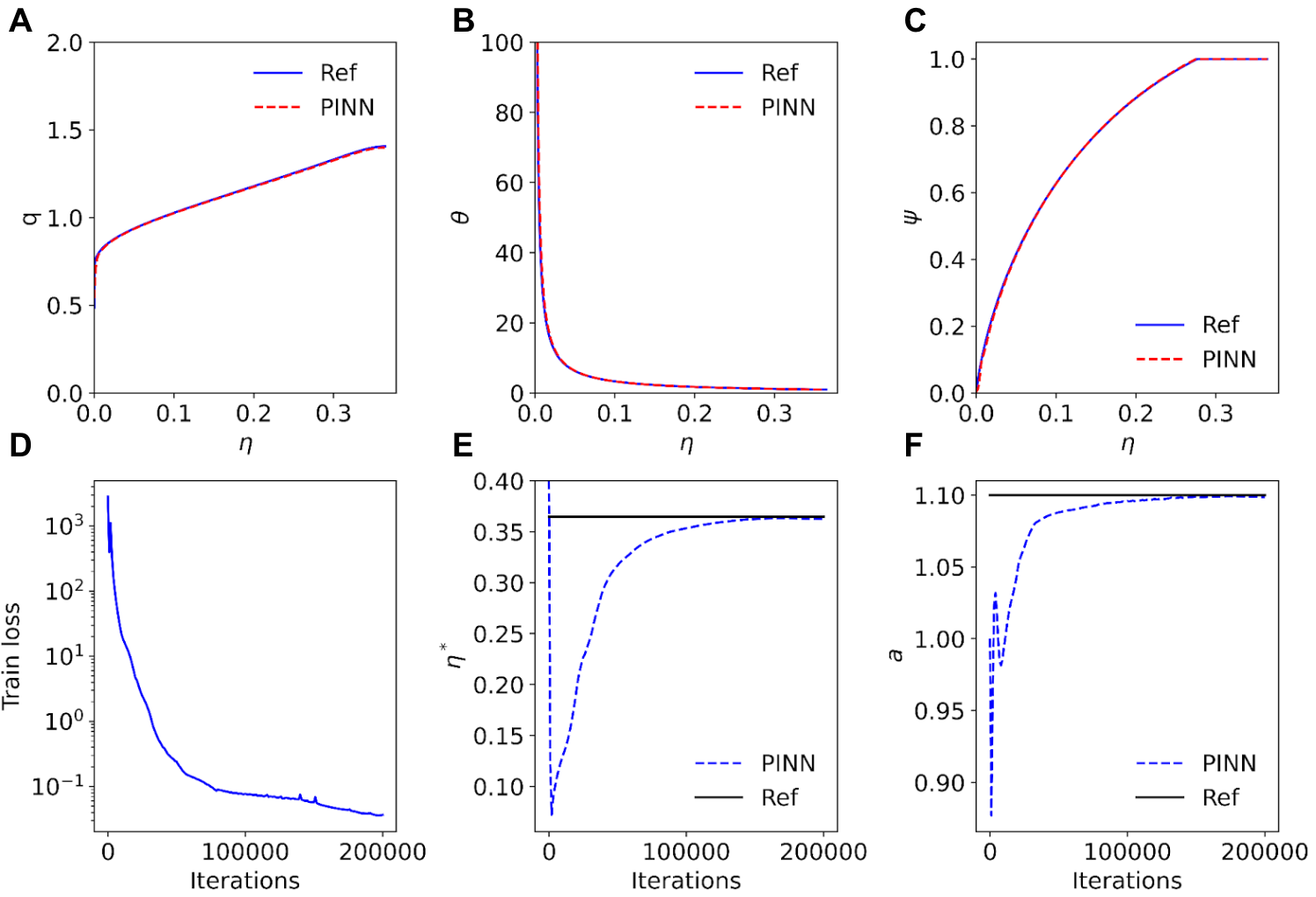}
    \caption{\textbf{PINN results for the inverse B\&S model with one unknown parameter.} (\textbf{A}) The reference value and prediction for $q$. (\textbf{B}) The reference value and prediction for $\theta$. (\textbf{C}) The reference value and prediction for $\psi$. (\textbf{D}) The value of train loss throughout training. (\textbf{E}) The convergence of $\eta^*$ throughout training. (\textbf{F}) The convergence of $a$ throughout training.}
    \label{fig:iap-inv-one-var}
\end{figure}

The detailed training results for a trial with initial guess $a=1$ are displayed in Fig.~\ref{fig:iap-inv-one-var}, while the results from ten trials with varying initial guesses of $a$ are displayed in Table~\ref{tab:bs_inv_results_one_var}. In Figs.~\ref{fig:iap-inv-one-var}A, B, and C, we see the reference solution and PINN predictions for $q$, $\theta$, and $\psi$ are nearly identical. We consistently achieve this in the ten random trials, as Table~\ref{tab:bs_inv_results_one_var} shows the $L^2$ relative errors of $q$, $\theta$, and $\psi$ are always well under $1\%$. In Fig.~\ref{fig:iap-inv-one-var}D, we see that the training loss decreases steadily, converging to a solution. In fact, the consistently accurate results in Table~\ref{tab:bs_inv_results_one_var} suggest the PINN consistently converges to the global minimum. In Figs.~\ref{fig:iap-inv-one-var}E and F, we find that the endogenous variable $\eta^*$ and unknown parameter $a$ quickly converge to the true values. In the ten trials, $a$ is always accurately estimated, with a final value of $a = 0.1099 \pm 0.0007$ (Table~\ref{tab:bs_inv_results_one_var}). Finally, Table~\ref{tab:bs_inv_results_one_var} suggests we were able to successfully match the moment value of $a_\mathrm{target}$, with the errors in our predictions always being well under $0.5\%$.

\begin{table}[htbp]
    \centering
    \caption{\textbf{Results from ten trials of the inverse B\&S model with one unknown parameter.} The initial value of $a$ is randomly sampled in the range $(0.06, 0.16)$.}
    \label{tab:bs_inv_results_one_var}
    \begin{tabular}{c|cc|cc|ccc}
    \toprule
    Trial & \multicolumn{2}{c}{$a$} & \multicolumn{2}{c}{$a_{\text{target}}$} & \multicolumn{3}{c}{$L^2$ relative error} \\
    & Initial & Final & Value & Error & $q$ & $\theta$ & $\psi$ \\
    \midrule
1 & 0.1500 & 0.1098 & 0.1093 & 0.16\% & 0.17\% & 0.53\% & 0.06\% \\
2 & 0.0921 & 0.1098 & 0.1093 & 0.22\% & 0.22\% & 0.69\% & 0.08\% \\
3 & 0.1472 & 0.1100 & 0.1095 & 0.01\% & 0.01\% & 0.02\% & 0.00\% \\
4 & 0.1230 & 0.1099 & 0.1094 & 0.11\% & 0.12\% & 0.36\% & 0.04\% \\
5 & 0.1236 & 0.1098 & 0.1094 & 0.14\% & 0.14\% & 0.45\% & 0.05\% \\
6 & 0.1005 & 0.1098 & 0.1093 & 0.22\% & 0.23\% & 0.71\% & 0.09\% \\
7 & 0.1509 & 0.1099 & 0.1094 & 0.08\% & 0.08\% & 0.26\% & 0.03\% \\
8 & 0.1415 & 0.1099 & 0.1094 & 0.07\% & 0.07\% & 0.21\% & 0.03\%  \\
9 & 0.0889 & 0.1099 & 0.1094 & 0.09\% & 0.09\% & 0.29\% & 0.03\%  \\
10 & 0.0645 & 0.1098 & 0.1093 & 0.15\% & 0.16\% & 0.49\% & 0.06\%  \\
\bottomrule
\end{tabular}
\end{table}

\subsubsection{Estimation of two parameters}
\label{sec:iap-two-params}

Next, we estimate two parameters ($a$ and $\sigma$) from the moment conditions:
\begin{align*}
a_{\text {target }} &= \int_{0}^{\eta^{*}}[\psi(\eta) a+(1-\psi(\eta)) \underline{a}] f(\eta) d \eta \quad \text{and} \quad \mathrm{Vol}_{\text {target }} = \frac{\sqrt{\int_{0}^{\eta^{*}}(q-\bar{q})^{2} f(\eta) d \eta}}{\bar{q}},
\end{align*}
where $\bar{q}$ is defined as $\int_{0}^{\eta^{*}}q(\eta) f(\eta) d \eta$,  $a_{\text {target }} = 0.1095$, and $\mathrm{Vol}_{\text{target}} = 0.04126$.

A large part of our implementation is the same as Section~\ref{sec:iap-one-param}. However, instead of explicitly solving for $\psi$, noting that the left hand side of Eq.~\eqref{eq:psi} is monotonic, we utilize a basic bisection algorithm to solve for $\psi$. In addition to scaling $a$ as in Section~\ref{sec:iap-one-param}, we also scale $\sigma$ up by $100$ and scale it back down after training. Our loss weights for this section are slightly different: $10^6$ for the ODE on $q''$, $2 \times 10^5$ for the ODE on $\theta''$, $10^6$ for $q$ increasing, $10^4$ for $\theta$ decreasing, $10^5$ for the $a_{\mathrm{target}}$ moment condition, $5 \times 10^3$ for the $\mathrm{Vol}_\mathrm{target}$ moment condition, and $10^1$ for the $q(0) = \underline{q}$ boundary condition.

\paragraph{Result of a successful trial.}
For the training of PINNs, the unknown parameters ($a$ and $\sigma$) are randomly initialized, and thus the final values of $a$ and $\sigma$ may depend on the initial values. In our previous problems, the final values are not sensitive to the initial values, but in this problem, we will show that the final values of $a$ and $\sigma$ are sensitive to their initial values.

\begin{figure}[htbp]
    \centering
    \includegraphics[width=\textwidth]{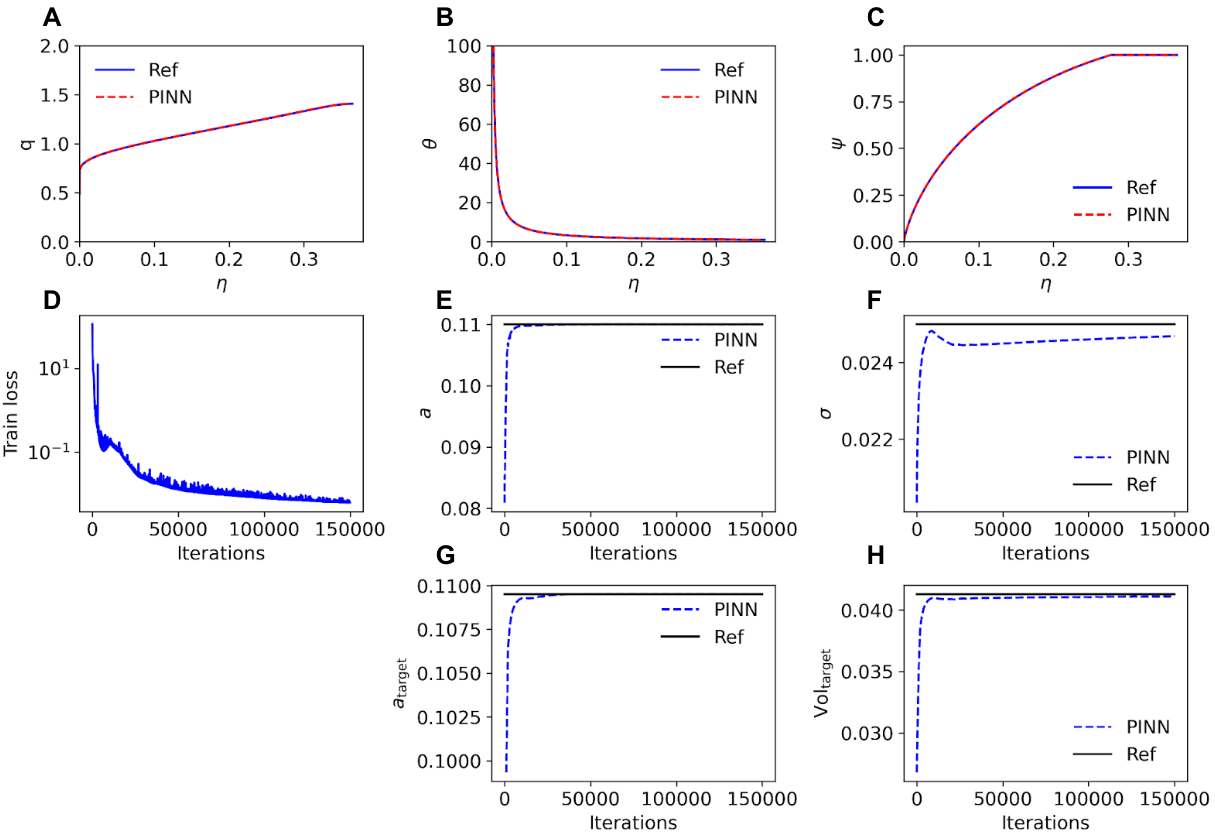}
    \caption{\textbf{PINN results for the inverse B\&S model with two unknown parameters.} (\textbf{A}) The reference value and prediction for $q$. (\textbf{B}) The reference value and prediction for $\theta$. (\textbf{C}) The reference value and prediction for $\psi$. (\textbf{D}) The value of train loss throughout training. (\textbf{E}) The convergence of $a$ throughout training. (\textbf{F}) The convergence of $\sigma$ throughout training. (\textbf{G}) The trajectory of $a_\mathrm{target}$ throughout training. (\textbf{H}) The trajectory of $\mathrm{Vol}_\mathrm{target}$ throughout training. The results here correspond to Trial 3 in Fig.~\ref{fig:bs_inv_param_trajectories}A and Table~\ref{tab:bs_inv_results_two_var}.}
    \label{fig:bs_inv_graphs}
\end{figure}

\begin{figure}[htbp]
    \centering
    \includegraphics[scale = 0.4]{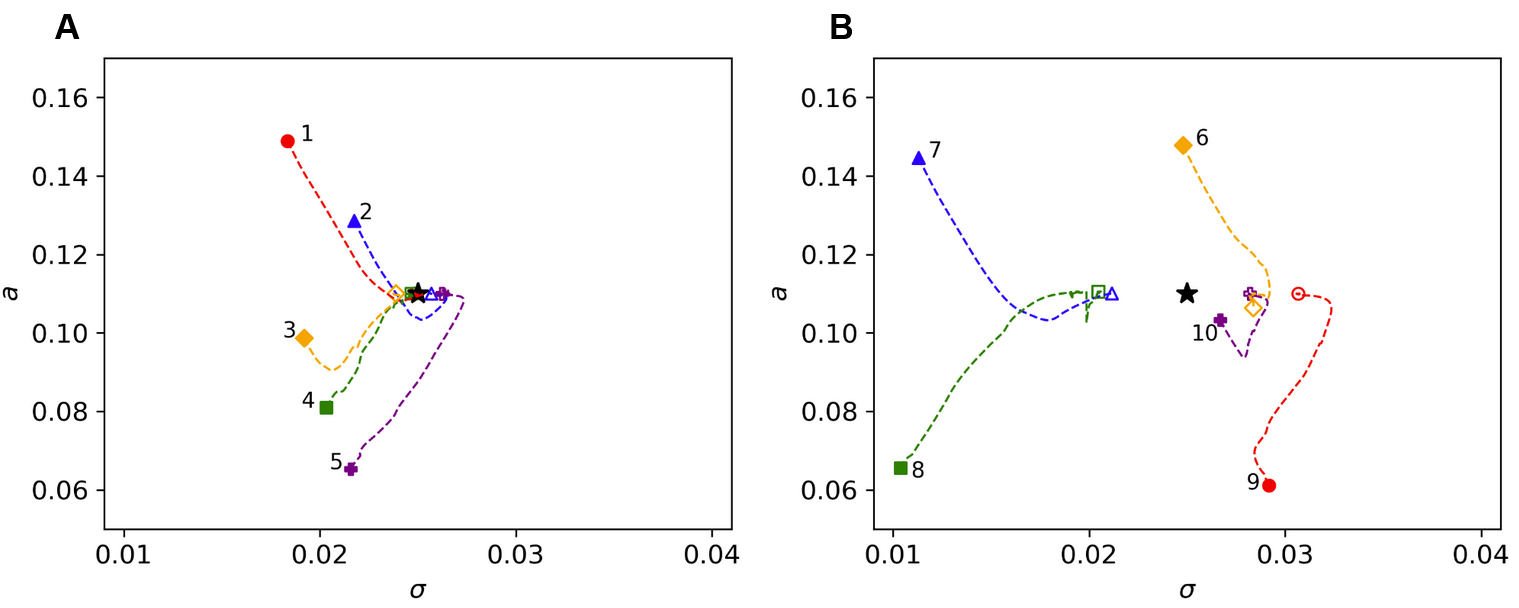}
    \caption{\textbf{Convergence trajectories in the parameter space $(\sigma, a)$.} (\textbf{A}) Five cases that converge to the true solutions. (\textbf{B}) Five cases that fail to converge to the true solutions. The black star in the center of the figure represents the true value $(\sigma, a)$, the shaded-in shape represents the initial guess, the empty shape represents the final estimation, and the path connecting them is the trajectory. Also see Table~\ref{tab:bs_inv_results_two_var} for more details.}
    \label{fig:bs_inv_param_trajectories}
\end{figure}

Here, we first show the results for an example with the initial guesses $a=0.0809$ and $\sigma=0.0203$ (Fig.~\ref{fig:bs_inv_graphs}). In this example, the training loss steadily decreases as the PINN converges to a solution (Fig.~\ref{fig:bs_inv_graphs}D). After training, the PINN predictions and reference solutions are nearly identical for $q$, $\theta$, and $\psi$ (Figs.~\ref{fig:bs_inv_graphs}A, B, and C). During training, the value of $a$ quickly converges to the reference value (Fig.~\ref{fig:bs_inv_graphs}E). The value of $\sigma$ slowly approaches the true value (Fig.~\ref{fig:bs_inv_graphs}F), though it is not as accurate as $a$. The convergence trajectory in the parameter space of $(\sigma, a)$ is also shown in Fig.~\ref{fig:bs_inv_param_trajectories}A, Trial 3. The errors of the final values of $a$ and $\sigma$ are $0.005\%$ and $1.24\%$, respectively, so both parameters had relatively accurate predictions. Moreover, we were able to successfully match both moment targets of $a_\mathrm{target}$ and $\mathrm{Vol}_\mathrm{target}$, as both moments quickly converged to their true values (Figs.~\ref{fig:bs_inv_graphs}G and H). The initial and final values of $a$ and $\sigma$, as well as the errors of the final $a_\mathrm{target}$, $\mathrm{Vol}_\mathrm{target}$, $q$, $\theta$, and $\psi$ are summarized in Table~\ref{tab:bs_inv_results_two_var}, Trial 3.

\begin{table}[htbp]
    \small
    \centering
    \caption{\textbf{PINN parameter estimation from different initial values of $a$ and $\sigma$.} Trials 1--5 converge to the true solutions. Trials 6--10 fail to converge to the true solutions. The trial numbers in the table correspond to those in Fig.~\ref{fig:bs_inv_param_trajectories}.}
    \label{tab:bs_inv_results_two_var}
    \begin{tabular}{c|cc|cc|cccc|ccc}
    \toprule
    Trial & \multicolumn{2}{c}{Initial values} & \multicolumn{2}{c}{Final values} & \multicolumn{2}{c}{$a_{\text{target}}$} & \multicolumn{2}{c}{Vol$_{\text{target}}$} & \multicolumn{3}{c}{$L^2$ relative errors} \\
    & $a$ & $\sigma$ & $a$ & $\sigma$ & Value & Error & Value & Error & $q$ & $\theta$ & $\psi$ \\
    \midrule
1 & 0.1489 & 0.0183 & 0.1100 & 0.0249 & 0.1095 & 0.00\% & 0.0412 & 0.19\% & 0.06\% & 0.10\% & 0.05\% \\
2 & 0.1286 & 0.0217 & 0.1100 & 0.0257 & 0.1095 & 0.01\% & 0.0416 & 0.88\% & 0.33\% & 0.56\% & 0.22\% \\
3 & 0.0809 & 0.0203 & 0.1100 & 0.0247 & 0.1095 & 0.01\% & 0.0411 & 0.37\% & 0.15\% & 0.26\% & 0.10\% \\
4 & 0.0987 & 0.0192 & 0.1100 & 0.0239 & 0.1095 & 0.01\% & 0.0407 & 1.43\% & 0.55\% & 0.87\% & 0.37\% \\
5 & 0.0653 & 0.0216 & 0.1099 & 0.0262 & 0.1094 & 0.07\% & 0.0419 & 1.47\% & 0.65\% & 1.20\% & 0.38\% \\
\midrule
6 & 0.0612 & 0.0292 & 0.1100 & 0.0307 & 0.1096 & 0.04\% & 0.0392 & 5.00\% & 1.96\% & 2.79\% & 1.31\% \\
7 & 0.1446 & 0.0113 & 0.1100 & 0.0212 & 0.1100 & 0.42\% & 0.0391 & 5.23\% & 2.71\% & 4.45\% & 1.41\% \\
8 & 0.0655 & 0.0104 & 0.1104 & 0.0205 & 0.1059 & 3.32\% & 0.0408 & 1.23\% & 4.71\% & 12.35\% & 0.35\% \\
9 & 0.1478 & 0.0248 & 0.1064 & 0.0284 & 0.1095 & 0.03\% & 0.0429 & 4.05\% & 1.50\% & 2.71\% & 1.03\% \\
10 & 0.1033 & 0.0267 & 0.1100 & 0.0282 & 0.1094 & 0.06\% & 0.0442 & 7.09\% & 2.60\% & 4.97\% & 1.79\% \\ \bottomrule
    \end{tabular}
\end{table}

\paragraph{Results of both successful and failed trials.}
In this problem, we find that the predicted values of $a$ and $\sigma$ are sensitive to the initial value of $\sigma$, but not to the initial value of $a$. For example, the trajectories of five trials that converge from different random initial values to the true solutions are displayed in Fig.~\ref{fig:bs_inv_param_trajectories}A, while the trajectories of five trials that fail to converge to the true solutions are displayed in Fig.~\ref{fig:bs_inv_param_trajectories}B. Furthermore, we find that the prediction of $a$ is consistently accurate, regardless of the initial values. For all the ten trials, the initial and final values of $a$ and $\sigma$, as well as the errors of the final $a_\mathrm{target}$, $\mathrm{Vol}_\mathrm{target}$, $q$, $\theta$, and $\psi$ are summarized in Table~\ref{tab:bs_inv_results_two_var}.

\paragraph{Error convergence using random initial values.}
As discussed above, there is no guarantee that the training with random initial values of $a$ and $\sigma$ would converge to the true solution, which seems to be an issue. However, trials that converge to a wrong solution can be easily identified, because $\mathrm{a}_\mathrm{target}$ and $\mathrm{Vol}_\mathrm{target}$ are farther from the true values (see Trials 6--10 in Table~\ref{tab:bs_inv_results_two_var}). Hence, we can use the errors of $\mathrm{a}_\mathrm{target}$ and $\mathrm{Vol}_\mathrm{target}$ to determine if the found solution is good or not. Based on this, we can simply run multiple trials and select the best solution (in terms of $\mathrm{a}_\mathrm{target}$ and $\mathrm{Vol}_\mathrm{target}$) as the final solution.

\begin{figure}[htbp]
    \centering
    \includegraphics[scale=0.45]{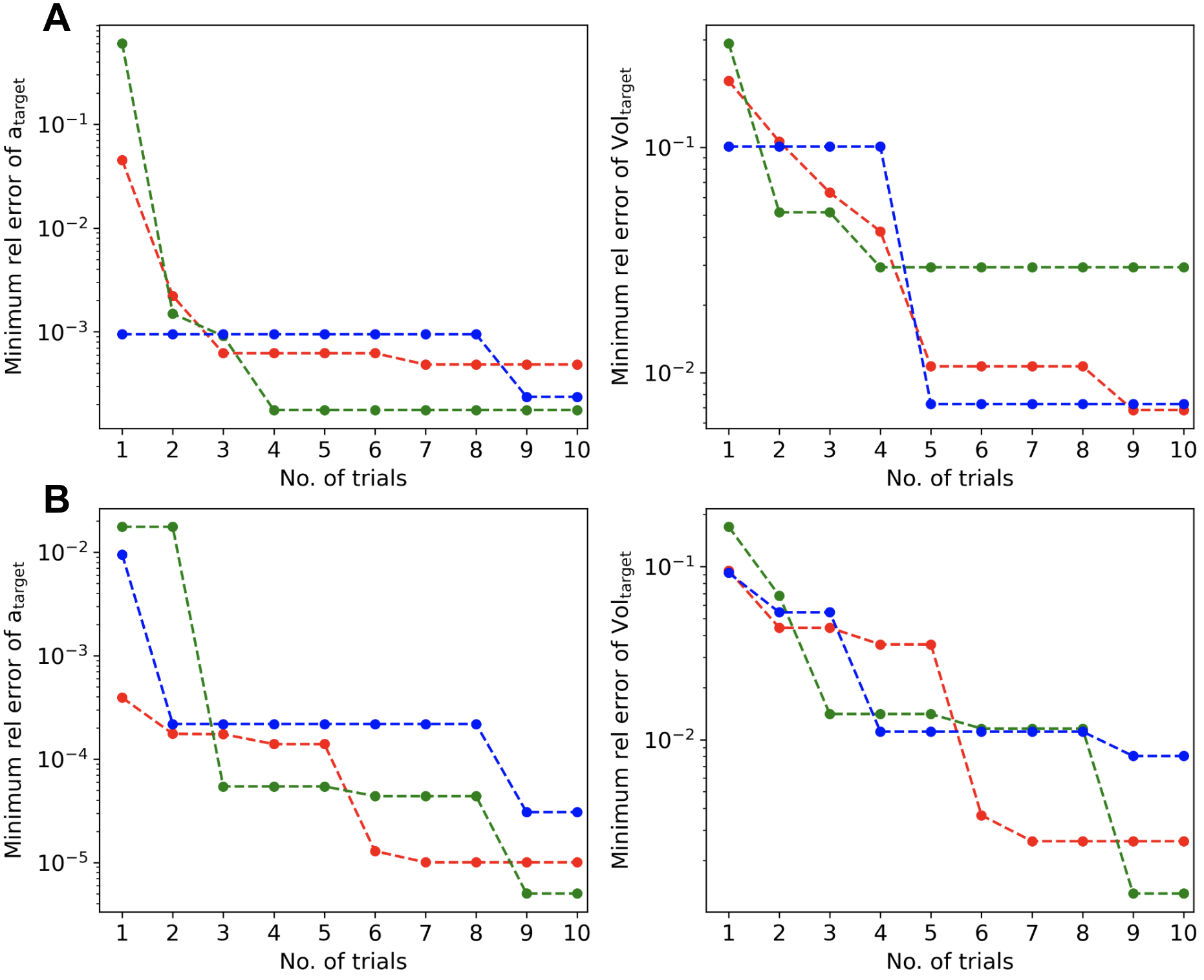}
    \caption{\textbf{Convergence of minimum relative errors of $a_\mathrm{target}$ and $\mathrm{Vol}_\mathrm{target}$ with respect to the number of trials using random guesses and Bayesian optimization.} (\textbf{A}) The initial guesses of $a$ and $\sigma$ are chosen randomly from the intervals $(0.06, 0.16)$ and $(0.01, 0.04)$, respectively. (\textbf{B}) The first five initial guesses of $a$ and $\sigma$ are chosen at random from the intervals $(0.06, 0.16)$ and $(0.01, 0.04)$, respectively. The five guesses afterwards are chosen using Bayesian optimization. In each figure, we show three independent runs, and each curve is the convergence of a single run.}
    \label{fig:iap_inv_trials}
\end{figure}

Specifically, we perform the first trial, where we select $a$ and $\sigma$ at random in the ranges $(0.6, 1.6)$ and $(0.01, 0.04)$, respectively, and we compute the final errors of $\mathrm{a}_\mathrm{target}$ and $\mathrm{Vol}_\mathrm{target}$. Then, we perform the second trial with different random initial values and compute the errors of $\mathrm{a}_\mathrm{target}$ and $\mathrm{Vol}_\mathrm{target}$. Now we have two sets of results from the first and second trials, and we select the one with the smaller error. We repeat this procedure until we have ten trials, and we always select the result with the smallest error. After ten trials, we obtain the convergence of the smallest errors of $\mathrm{a}_\mathrm{target}$ and $\mathrm{Vol}_\mathrm{target}$ with respect to the number of trials, and examples of three runs are shown in Fig.~\ref{fig:iap_inv_trials}A. Although each trial is random and cannot guarantee a good solution, we generally receive satisfactory results after ten trials. In particular, the relative error of $a_\mathrm{target}$ is consistently under $10^{-3}$, and the minimum relative error of $\mathrm{Vol}_\mathrm{target}$ generally hovers around $10^{-2}$.

\paragraph{Error convergence using Bayesian optimization.}
To further improve the convergence rate, instead of using random initial guesses for each trial, Bayesian optimization \citep{snoek2012bayesian} can be applied. Bayesian optimization is a method to find the maximum or minimum value of a target function given a small number of function values. In particular, it uses Bayes' theorem to develop a probabilistic model of the function, and then Bayesian optimization selects the next point on the function with the highest possible value. In our problem, the target function has initial guesses $a$ and $\sigma$ as the parameters, and the function value is the errors of $\mathrm{a}_\mathrm{target}$ and $\mathrm{Vol}_\mathrm{target}$ after training.

Specifically, we first run five random trials, and then we apply Bayesian optimization to determine the initial values of the 6th trial. After we have the result of the 6th trial, we use all the first six trials to determine the initial values of 7th trial. We repeat this procedure until we have ten trials. We find that Bayesian optimization produces significantly faster convergence and better results than random trials (Fig.~\ref{fig:iap_inv_trials}B). For both $a_\mathrm{target}$ and $\mathrm{Vol}_\mathrm{target}$, the minimum relative error with Bayesian optimization is about an order of magnitude lower than that of random initial values.
\section{Conclusions}
\label{sec:conclusion}

In this study, we proposed a new method for solving economics models with deep learning through physics-informed neural networks (PINNs). We demonstrated the advantages of this method: generality, simultaneous solution and estimation, leveraging the state-of-art machine-learning techniques, and handling large state space. Our proposed method provides a general framework for solving partial differential equations (PDEs), whereas in traditional methods, we usually need to design different algorithms for different problems. Furthermore, the inverse problem can be solved very easily and is no more than adding an extra loss term for additional information. We showed the effectiveness of the PINN method in solving two models (industrial dynamics with financial frictions and a macroeconomic model with the financial sector), each featuring its own challenges.

There are a couple of directions for future research. First, currently, we only allow for heterogeneous agents with stationary population distribution. Our method can be generalized to deal with mean-field games where the distribution is a state variable that changes over time. Second, we have only solved exactly identified estimation problems where the number of moment conditions is equal to the number of parameters to be estimated. A more generalized method should be able to handle overidentified systems and provide error bounds on the estimated coefficients, similar to the classical generalized method of moments (GMM) estimations. Finally, our method can be applied to a broader variety of economics and finance problems.


\section*{Acknowledgements}
B.F., E.Q., and L.L thank MIT's PRIMES-USA program.

\appendix
\section{Firm industrial dynamic model}
\label{secApx:firm dynamic}

\begin{proposition}
In bank industrial dynamic models, stationary distribution exists.
\end{proposition}
\begin{proof}
    As the volatility $\sigma(z)$ is constant, and Ornstein–Uhlenbeck process is recurrent, we only need to check if the linear growth condition for $(\mu_e,\mu_z)$ is satisfied:
    \[
        \begin{aligned}
            ||\vec{\mu}(e,z)|| \leq ||\mu_e(e,z)|| + ||\mu_z(e,z)|| &= ||\theta_0 (z-z_m)|| + ||\pi^*(e,z) - \textbf{1}_{(e_t,z_t)\in\mathscr{C}^c}\kappa(\phi e-f(z,l^*))^+||\\
            & \leq ||\theta_0 (z-z_m)|| + ||\pi^*(e,z)|| + ||\kappa(\phi e-f(z,l^*))^+||\\
            & \leq ||\theta_0 (z-z_m)|| + ||\pi^*(e,z)|| + |\kappa|(||(\phi e)||+||f(z,l^*)||)\\
            & \leq  ||\theta_0 (z-z_m)|| + ||\pi^*(e,z)|| + |\kappa|(|\phi||x|+||f(z,l^*)||).
        \end{aligned}
    \]
    We can conclude our proof by noting that both instantaneous profit $\pi^*(e,z)$ and deposit production $f(z,l^*)$ are scaled by productivity $z$.
\end{proof}

\section{Macroeconomic model with a financial sector}
\label{secApx:MacroFin}

The model in \citep{brunnermeier2014macroeconomic} considers an economy populated with one unit of experts (indexed by $i,i\in\mathbb{I}=[0,1]$) and one unit of households (indexed by $j,j\in \mathbb{J}=[1,2]$), in an infinite horizon setup. Time is continuous here. The physical capital $k_t$ held by experts produces output at rate $y_t=ak_t$, while households produces at rate $\underline{y}_t = \underline{a}k_t$. Denoting the investment rate per unit of capital as $\iota_t,\underline{\iota}_t$ for experts and households respectively (i.e., $\iota_tk_t$ is the total investment rate of experts), the corresponding capital evolves as
\[
    \left\{
    \begin{gathered}
        dk_t = (\Phi(\iota_t)-\delta)k_t dt +\sigma k_t dZ_t,\\
        d\underline{k}_t = (\Phi(\underline{\iota}_t)-\underline{\delta})\underline{k}_tdt+\sigma\underline{k}_tdZ_t,
    \end{gathered}\right.
\]
where $Z_t$ here is exogenous aggregate Brownian shocks, $\Phi(\cdot)$ is standard investment technology with convex adjustment costs, i.e., $\Phi(0) = 0,\Phi'(0)=1,\Phi'(\cdot)>0$ and $\Phi''(\cdot)<0$, $\delta,\underline{\delta}$ are depreciation rate (assume $\underline{\delta}>\delta$), and $\sigma$ is the volatility. Experts and households are risk neutral, households have discount rate $r$ and they may have positive or negative consumption, which means households provide fully elastic lending at risk-free rate $r$. 
The equilibrium market price of capital is postulated as a Geometric Brownian Motion:
\[
    dq_t = \mu_{t}^q q_t dt + \sigma_t^q q_t dZ_t,
\]
where $\mu_t$ can be viewed as the time dependent price drift and $\sigma_t^q$ is the price volatility. Both $\mu^q_t$ and $\sigma^q_t$ are determined by market equilibrium. We focus on the case that equilibrium price $q\in[\underline{q},\overline{q}]$, where $\underline{q} = \max_{\iota}\frac{\underline{a}-\iota}{r-(\Phi(\iota)-\underline{\delta})}$ and $q = \max_{\iota}\frac{a-\iota}{r-(\Phi(\iota)-\delta)}$. The return of the capital managed by experts and households can then be expressed by
\[
    \begin{gathered}
        dr_t^k = \frac{dD_t+d(k_tq_t)}{k_tq_t} =\frac{(a-\iota_t){k_t}}{{k_t} q_t} + \left(\Phi(\iota_t)-\delta+\mu_t^q+\sigma\sigma_t^q\right)dt+(\sigma+\sigma_t^q)dZ_t,\\
        d\underline{r}_t^k = \frac{d\underline{D}_t+d(k_tq_t)}{k_tq_t} =\frac{(\underline{a}-\underline{\iota}_t){k_t}}{{k_t} q_t} + \left(\Phi(\underline{\iota}_t)-\underline{\delta}+\mu_t^q+\sigma\sigma_t^q\right)dt+(\sigma+\sigma_t^q)dZ_t.
    \end{gathered}
\]
The cumulative consumption of a household and an expert are denoted as $\underline{c}_t$ and $c_t$, respectively. Then utilities are given by
\[
\mathbb{E}\left[\int_0^\infty e^{-rt}d\underline{c}_t\right] \text{ (households) } \quad \text{and} \quad
\mathbb{E}\left[\int_0^\infty e^{-\rho t}dc_t\right] \text{ (experts)}.
\]
The net worth of an expert  $n_t$ evolves as
\[
    \frac{dn_t}{n_t} = x_t dr_t^k + (1-x_t)rdt-\frac{dc_t}{n_t},
\]
where $x_t$ is the fraction of capital. The first part is return on risky assets, the second part is return on safe assets, and the third part is consumption. Similarly, for households, we have
\[
    \frac{d\underline{n}_t}{\underline{n}_t} = \underline{x}_t d\underline{r}_t^k + (1-\underline{x}_t)rdt-\frac{d\underline{c}_t}{\underline{n}_t},
\]
where $\underline{x}_t$ is the fraction of capital.
Both households and experts maximize their utility. Households can have negative consumption while experts cannot, i.e., $dc_t\geq 0$. Households and experts' problems can be written as
\[
    \max_{\underline{x}_t\geq0,d\underline{c}_t,\underline{\iota}_{t}}\mathbb{E}\left[\int_0^\infty e^{-rt}d\underline{c}_t\right] \text{ (households), }
    \max_{x_t\geq0,dc_t\geq 0,\iota_{t}}\mathbb{E}\left[\int_0^\infty e^{-\rho t}dc_t\right] \text{ (experts)},
\]
subject to net worth's equation of motion.\par

\paragraph{Definition for equilibrium.}
{\it Given initial wealth distribution, $k_0^i,k_0^j$, an equilibrium is described by the stochastic process $\{q_t,n_t^i,\underline{n}_t^i\geq 0,n_t^i,\underline{k}_t^i\geq 0, \iota_t^i,\underline{\iota}_t^i,dc_t^i\geq 0, \underline{dc}_t^i\}$, such that: (1) initial net worth is $n_0^i=k_0^iq_0$, $\underline{n}_0^j = k_0^j q_0$; (2) each expert and agent solves their problems, given capital price $q_t$; (3) markets for consumption goods and capital are clear, i.e.,
\[
\begin{gathered}
    \int_{\mathbb{I}} dc_t^i di+\int_{\mathbb{J}} d\underline{c}_t^jdj = \int_{\mathbb{I}} (a-\iota_t^i)k_t^i di+\int_{\mathbb{J}} (\underline{a}-\underline{\iota}_t^j)\underline{k}_t^jdj, \\
    \int_{\mathbb{I}}k_t^i di + \int_{\mathbb{J}}\underline{k}_t^j dj = K_t, \text{$dK_t$ is: }\left(\int_{\mathbb{I}}(\Phi(\iota_t^i))-\delta)k_t^i di + \int_{\mathbb{J}}(\Phi(\underline{\iota}_t^i))-\underline{\delta})\underline{k}_t^j dj\right)dt + \sigma K_t dZ_t.
\end{gathered}
\]}

\paragraph{Solution.}
First, households and experts' investment choices $\iota_t,\underline{\iota}_t$ are solved by maximizing $dr_t^k,d\underline{r}_t^k$, which means
\[
    \iota_t,\underline{\iota}_t\in \arg \max_{\iota}\Phi(\iota) - \iota/q_t\Rightarrow \iota_t=\underline{\iota}_t={\Phi'}^{-1}(1/q_t). 
\]
Second, denoting $\psi_t$ as the fraction of capital held by experts, we are led to equilibrium condition $\mathbb{E}_{t}[d\underline{r}^k_t]/dt\leq r$, with equality if $1-\psi_t> 0$. This condition means that when the expected return is less than the risk-free asset, households will not hold any capital. Also, risk-neutral households will hold a fraction of capital when the return equals the risk-free rate.

Third, to solve the experts' problem, we introduce multiplier $\theta_t$ for experts' future utility, i.e., $\theta_t n_t \equiv \mathbb{E}_t\left[\int_0^\infty e^{-\rho(s-t)}dc_s\right]$. Introducing unit worth's consumption $d\zeta_t$, the experts optimal trading strategy is
\[
    \begin{aligned}
        \rho \theta_t n_t &=\max_{\hat{x}_t,d\zeta_t} n_t d\zeta_t + \mathbb{E}d[\theta_t n_t]\\
    \end{aligned}
\]
Considering a finite process $d\theta_t/\theta_t = \mu_{t}^{\theta} dt + \sigma_{t}^{\theta}dZ_{t}$, for optimal strategy, the solution of $\theta_t$ features: (1) it is always true that $\theta_t\geq 1$, $d\zeta_t>0$ only when $\theta_t =1$; (2) $\mu_t=\rho - r$; (3) either $x_t>0$ when $\mathbb{E}_t[dr_t^k]/dt-r = -\sigma_t^\theta (\sigma+\sigma_t^q)$ (the risk premium), or $x_t=0$ when $\mathbb{E}_t[dr_t^k]/dt-r < -\sigma_t^\theta (\sigma+\sigma_t^q)$. \par

Denote the experts' wealth share as $\eta_t\equiv \frac{N_t}{q_tK_t}\in[0,1]$, where all functions can be functions of $\eta_t$. By applying It\^{o}'s lemma, we get ($\left<\cdot,\cdot\right>$ is the quadratic variation)
\[
    \begin{aligned}
        d\eta_t &= \frac{dN_t}{q_t K_t} + N_t d\left(\frac{1}{q_tK_t}\right)+\left<dN_t,d\left(\frac{1}{q_tK_t}\right)\right>\equiv\eta_t(\mu_t^\eta dt +\sigma_t^\eta dZ_t)\\
        \Rightarrow \mu_t^\eta &= - \sigma_t^\eta(\sigma +\sigma_t^q +\sigma_t^\theta)+ \frac{a-\iota(q_t)}{q_t} + (1-\psi_t)(\underline{\delta}-\delta),\ 
        \sigma_t^\eta = \frac{\psi_t-\eta_t}{\eta_t}(\sigma+\sigma_t^q).
    \end{aligned}
\]

\paragraph{Equilibrium conditions.}
Optimal strategies of households and experts imply
\[
    \frac{\mathbb{E}_t[dr_t^k-d\underline{r}_t^k]}{dt}-\sigma_t^\theta(\sigma+\sigma_t^q)=0\Rightarrow \psi(\eta),
\]
which means for experts, the opportunity cost of holding capital is the risk premium, if the equilibrium has an interior solution. When the implied solution $\psi>1$, the above equation does not necessarily hold, as it is always profit profitable to hold capital in this case. Next, by It\^{o}'s formula, we can solve the equilibrium price $q(\eta)$ and multiplier $\theta_t$ from
\[
    \left\{
    \begin{array}{c}
        \mu_t^q q(\eta) = q'(\eta)\mu_{t}^\eta \eta + \frac{1}{2}(\sigma_t^\eta)^2 \eta^2 q''(\eta), \\
        \mu_t^\theta\theta(\eta) = \theta'(\eta)\mu_{t}^\eta \eta + \frac{1}{2}(\sigma_t^\eta)^2 \eta^2 \theta''(\eta).
    \end{array}\right.
\]
Given the stochastic process of $d\eta_t = \eta_t\mu_t^\eta dt + \eta_t\sigma_t^\eta$, the distribution $f(\eta,t)$ evolves as
\[
    \frac{\partial}{\partial t} f(\eta,t) =-\frac{\partial}{\partial \eta}\left(\mu^\eta f(\eta,t)\right) + \frac{1}{2}\frac{\partial^2}{\partial \eta^2}\left(\sigma^2_\eta(\eta)f(\eta,t)\right).
\]
Prior to solving the invariant distribution, we need to show that the distribution function exists. For example, consider a Geometric Brownian Motion: $dX_t = \mu X_t dt+\sigma X_t dZ_t$ with reflecting boundary at $0,D$, the stationary distribution solved from Appendix~\ref{secApx:Dimension Reduction Trick} is: $f(x) = \frac{\frac{2\mu}{\sigma^2}-1}{D^{\frac{2\mu}{\sigma^2}-1}}x^{\frac{2\mu}{\sigma^2}-2}\times\textbf{1}_{x\in[0,D]}$, we can see that it cannot be an invariant distribution when $\frac{2\mu}{\sigma^2}-1<0$, as density is negative.

\begin{proposition}
In \citep{brunnermeier2014macroeconomic}, the stationary distribution exists, if $2(\rho-r)\sigma^2<\Lambda^2,\Lambda =  \frac{\mathbb{E}_t[dr_t^k-d\underline{r}_t^k]}{dt}=\frac{a-\underline{a}}{\underline{q}}-(\delta-\underline{\delta})$.
\end{proposition}
\begin{proof}
Stochastic process $\eta_t$'s recurrence is equivalent to the inequality, $\mu^\eta>\frac{(\sigma^\eta)^2}{2}$, when $\eta=0^+$ (see one dimensional case in \citep{risken1996fokker}). Asymptotic ansatz when $\eta\rightarrow0$ (in the online appendix of \citep{brunnermeier2014macroeconomic}): $\mu_{t}^\eta = \hat{\mu}+ o(1),\sigma_{t}^{\eta} = \hat{\sigma}+o(1),\psi(\eta)=C_\psi \eta+o(\eta) ,q(\eta)=\underline{q} + C_q \eta^\alpha + o(\eta^\alpha),\theta(\eta) = C_\theta\eta^{-\beta}+o(\eta^{-\beta})$ ($\alpha,\beta> 0$). By plugging into equilibrium condition, we have $\hat{\sigma}=\Lambda/\beta\sigma$. From the equations for $q(\eta)$ and $\theta(\eta)$, we find
\[
2\frac{(\rho-r)\beta^2\sigma^2}{\Lambda^2} = -\beta\frac{2\hat{\mu}}{\hat{\sigma}^2}+\beta(\beta+1)
\rightarrow \frac{2\hat{\mu}}{\hat{\sigma}^2}=\beta +1-\frac{2(\rho-r)}{\Lambda^2}\sigma^2 \beta.
\]
\end{proof}


\section{Order reduction in the forward equation}
\label{secApx:Dimension Reduction Trick}

This section is a technical note. We first discuss the structure of Kolmogorov forward equation (KFE), and then give a continuity equation's interpretation of it. Denote the density function as $f$; the KFE can be heuristically written as
\[
    \frac{\partial}{\partial t} f=-\sum_{i} \frac{\partial}{\partial x_{i}}\left(\mu_{i}(x) f\right)+\sum_{i, j} \frac{\partial^{2}}{\partial x_{i} \partial x_{j}}\left(\left(\sigma^{2}\right)_{i j}(x) f\right)\equiv \widehat{L}^{*} f=-\nabla \cdot \vec{J},
\]
where $\hat{L}^*$ is the Kolmogorov forward operator and the \textit{density flux} $\vec{J}$ is defined as
\[
    J_i = \frac{\partial}{\partial x_{i}}\left(\mu_{i}(x) f\right)-\sum_{j}\frac{\partial^{2}}{\partial x_{i} \partial x_{j}}\left(\left(\sigma^{2}\right)_{i j}(x) f\right), \quad \text{for the $i$--th column of $\vec{J}$}.
\]
To study the stationary distribution we are interested in, we assume the existence of stationary distribution, which means the following assumption holds.

\begin{assumption}
Regularity assumptions in \citep[Ch. 5]{karatzas1998brownian}.
\begin{enumerate} 
        \item $\sigma^2(x)$ is uniform elliptic, i.e.,
    \[
        \vec{y}^T[\sigma^2(x)]\vec{y}\geq \alpha |\vec{y}|^2,\ \forall \vec{y}\in\mathbb{R}^n.
    \]
    \item Coefficients are smooth and satisfy linear growth conditions
    \[
        \exists M\in\mathbb{R},\ s.t.\ ||\sigma^2(x)||\leq M,\ ||a(x)||\leq M(1+||x||),\ ||b(x)||\leq M(1+||x||),
    \]
    where $a(x),b(x)$ are defined as
    \[
        \begin{gathered}
            a_j(x)=-\mu_j(x)+\sum_{i=1}^{n}\partial_{ x_i}[\sigma^2(x)]_{ij}, \\
            b_j(x)=\frac{1}{2}\sum_{i,k}\partial^2_{x_i,x_j}[\sigma^2(x)]_{i,k}-\sum_{i}\partial_{x_i}\mu_{i}(x).
        \end{gathered}
    \]
    \item The stochastic process $X_t$ is recurrent.
    \end{enumerate}
\end{assumption}

For the reflecting boundary\footnote{Indifference condition for value function at the boundary implies the reflecting boundary for distribution. Usually, three boundary conditions are considered. They are: (1) \textit{refecting boundary} $\vec{J}\cdot\hat{n}|_{\partial\Omega}= 0$; (2) \textit{absorbing boundary}: $f|_{\partial \Omega} = 0$; (3) \textit{periodic boundary}: $\vec{J}|_{x = a} = \vec{J}|_{x=b}$.} in our problem, we have $\vec{J}\cdot\hat{n}\equiv 0$ at $\partial \Omega$. The \textbf{solution determination problem} can be formally written as
\[
\left\{
\begin{array}{c}
     \nabla \cdot\Vec{J} = 0,  \\
     \Vec{J}\cdot \hat{n}|_{\partial \Omega }=0,\\
     J_i = - \mu_i(x)f +\sum_j \frac{1}{2}\frac{\partial}{\partial x_j}((\sigma^2(x))_{ij}f),
\end{array}\right.
\]
with normalization condition $\int _\Omega fdV = 1$. 
\begin{proposition}
Under Assumption 1, the solution determination problem is equivalent to the problem
\[
    \vec{J}=0 \text{ and } \int_{\Omega}fdV = 1.
\]
\end{proposition}
\begin{proof}
    We first show that in the one-dimensional case (also see \citep{gabaix2016dynamics}), the probability flux is always zero under the reflecting boundary condition. This is because
    \[
        J(x) \equiv -\mu(x)f(x)+\frac{1}{2}\frac{d}{dx}\left(\sigma^2(x)f(x)\right)=\int_{\underline{x}}^{x}\hat{L}^*f(x')dx'+J(\underline{x}) =0,
    \]
    where $\underline{x}$ is the lower boundary.

    In higher dimensional cases, $\sigma^2(x)$ is the covariance matrix and $\vec{\mu}$ is the drift vector. According to \citep{pavliotis2014stochastic}, define  $Q(x)$ as: $\vec{Q}(x) = \left(\sigma^2(x)\right)^{-1}(2\vec{\mu}(x)-\nabla\sigma^2(x))$, where $\nabla\sigma^2(x)$ is defined as $\sum_{i,j}\frac{\partial}{\partial x_j}\sigma^2(x)_{ij}\vec{e}_i$. Then\footnote{This condition implies that the path integral exists, or $d\vec{Q}$ is integrable. Intuitively, for a constant $\sigma^2(x)$, this condition means that the drift term is \textit{curl-free}. Accordingly, a counter example in 2D can be constructed as: $\vec{\mu} = \frac{A}{r}\hat{\tau}$, where $\hat{\tau} = \frac{-y}{\sqrt{x^2+y^2}}\vec{e}_x + \frac{x}{\sqrt{x^2+y^2}}\vec{e}_y$.} $\vec{J}\equiv0$ \textbf{if and only if} $\frac{\partial Q_j}{\partial x_i}=\frac{\partial Q_i}{\partial x_j}$, for all $x\in\Omega$.  The density $f$ can be solved as
\[
    \begin{gathered}
        f(\vec{x}) = A\exp\left(-\frac{1}{2}\int_{x_0\in\partial\Omega}^{\vec{x}} \vec{Q}\cdot d\vec{x}\right) \quad \text{and} \quad \frac{1}{A} = \int_{\Omega}\exp \left(-\int_{x_0\in\partial\Omega}^x\vec{Q}(\vec{x}')\cdot d\vec{x}'\right)d^n\pmb{x}.
    \end{gathered}
\]
The above formula implies that once we obtain the solution for the KFE with reflecting boundary, we can conclude the solution is \textbf{unique} if Assumption 1 holds.
\end{proof}

\bibliographystyle{apalike}
\bibliography{main}

\section*{Statements and Declarations}

\paragraph{Funding.}
The authors declare that no funds, grants, or other support were received during the preparation of this manuscript.

\paragraph{Competing Interests.}
The authors have no relevant financial or non-financial interests to disclose.

\end{document}